%% file: newrecursion.tex
\documentclass[letterpaper,hyper]{JHEP3}
\usepackage[centertags]{amsmath}
\usepackage{wrapfig}
\usepackage{array,multirow}
\usepackage{amssymb}
\usepackage{graphicx}
\usepackage{color}
\usepackage{mathtools}
\usepackage{wasysym}
\usepackage[punctsep]{collref}
\collectsep[]{;}

\usepackage{epsfig}

\def\la{\lambda}

\def\lb{\bar{\lambda}}

\def\ep{\epsilon}

\def\Or[#1]{{\text{O}}\left({#1}\right)}
\def\dotl[#1,#2]{\left\langle #1,\, #2 \right\rangle}
\def\dotlb[#1,#2]{\left\langle #1,\, #2 \right\rangle}
\def\dotlm[#1,#2]{\left[ #1,\, #2 \right]}
\def\dotp[#1,#2]{(\vect{#1} \cdot\vect{#2})}
\def\aff[#1,#2]{\hat{#1}(#2)}
\def\n4sym{{\cal N}=4 SYM}
\def\>{\rangle}
\def\<{\langle}
\def\weight[#1,#2,#3]{\{(#1),#2,#3\}}
\def\ads[#1]{$\text{AdS}_{#1}$}

\newcommand{\be}{\begin{equation}}
\newcommand{\ee}{\end{equation}}
\newcommand{\ba}{\begin{align}}
\newcommand{\ea}{\end{align}}
\newcommand{\bs}{\begin{split}}
\def\sess\end{split}
\newcommand{\vect}[1]{{\boldsymbol{#1}}}
\newcommand{\norm}[1]{|{\boldsymbol{#1}}|}
\title{New Recursion Relations and a Flat Space Limit for AdS/CFT Correlators}
\author{Suvrat Raju \\ 
Harish-Chandra Research Institute, \\ Chatnag Road, Jhunsi, \\ Allahabad 211019.}
\abstract{
We consider correlation functions of the stress-tensor or a conserved current in AdS$_{d+1}$/CFT$_d$ computed using the Hilbert or the Yang-Mills action in the bulk. We introduce new recursion relations to compute these correlators at tree level.   These relations have an advantage over the BCFW-like relations described in  arXiv:1102.4724 and arXiv:1011.0780 because they can be used in all dimensions including $d=3$. We also introduce a new method of extracting flat-space S-matrix elements from AdS/CFT correlators in momentum space.  
We show that the $(d+1)$-dimensional flat-space amplitude of gravitons or gluons can be obtained as the
coefficient of a  particular singularity of the $d$-dimensional correlator of the stress-tensor or a conserved current; this technique is valid even at loop-level in the bulk. Finally, we show that our recursion relations automatically generate correlators that are consistent with this observation: they have the expected singularity and the flat-space gluon or graviton amplitude appears as its coefficient.
}
\preprint{HRI/ST/1201}
\begin{document}
\tableofcontents

\section{Introduction}

In this paper, we address two issues: the question of computing AdS/CFT correlators efficiently and the problem of reconstructing the flat-space S-matrix from boundary correlation functions. 

Given a perturbative bulk quantum field theory, the AdS/CFT conjecture \cite{Maldacena:1997re} 
provides a conceptually straightforward method 
of computing correlation functions in the boundary CFT \cite{Gubser:1998bc,Witten:1998qj}. However, in practice this procedure is quite tedious
for theories that involve gravitational interactions in the bulk. This is because of two difficulties. First, it is very difficult to compute graviton scattering amplitudes even in flat space since expanding the Hilbert action leads to an infinite set of interaction vertices of formidable complexity; for example, the four point vertex has 2,850 terms \cite{DeWitt:1967uc}. In AdS, this difficulty is compounded by the necessity of doing bulk integrals that, in position space, cannot be done in terms of elementary functions. 

In flat space, it was realized long ago \cite{DeWitt:1967uc,Parke:1986gb,Berends:1988zp} that complicated interaction vertices could nevertheless give rise to simple final answers for graviton amplitudes.
More recently, starting with the development of the BCFW recursion relations
\cite{Britto:2004ap,Britto:2005fq}, 
there has been rapid progress in the development of new on-shell techniques to compute amplitudes without using Feynman diagrams at all. (See \cite{Witten:2003nn,ArkaniHamed:2008gz,ArkaniHamed:2009dn,Forde:2007mi,Mason:2008jy, Spradlin:2008bu, Raju:2009yx,Lal:2009gn, Nguyen:2009jk,Lal:2010qq, Boels:2010nw} and references there.)  

In \cite{Raju:2010by, Raju:2011ed}, a generalization of the the BCFW recursion relations was presented that could be used to compute correlation functions of stress-tensors or conserved currents in AdS/CFT. The problem of performing difficult $z$-integrals was addressed in \cite{Maldacena:2011nz} which
made the observation that going to momentum space on the boundary led to simple answers for stress tensor correlators in odd boundary dimensions. (See also \cite{Fitzpatrick:2011ia,Paulos:2011ie,Penedones:2010ue} for a different approach to this problem.) 
 
However, these two results could not be immediately combined because although the BCFW recursion relations of \cite{Raju:2010by,Raju:2011ed} are phrased in momentum space, they apply only in higher than three boundary dimensions, while this is exactly the case that was considered in detail in \cite{Maldacena:2011nz}. 

In this paper we present new 
recursion relations for AdS$_{d+1}$/CFT$_d$ correlators in momentum space that are 
valid in arbitrary dimensions including, crucially,  $d=3$. Combined with the 
results for three-point functions presented in \cite{Maldacena:2011nz}, 
they can be used to compute explicit results for four-point functions
of the stress-tensor; we present these in a companion paper \cite{Raju:2012zs}. 

These recursion relations are somewhat similar to the recursion relations developed by Risager \cite{Risager:2005vk} for flat space gluon and graviton amplitudes. The idea is to shift the momentum of each operator by a vector that is proportional to the external polarization vector for that operator and a complex parameter $w$. This is very natural in $d=3$, where the polarization vectors of anything higher than a 3-pt correlator must be linearly dependent. Moreover, the behaviour at large $w$ is now fixed just by the Ward identities and does not require any additional analysis. This immediately leads to new recursion relations for flat-space gluon and graviton scattering amplitudes. Using the techniques of \cite{Raju:2010by,Raju:2011ed}, we can lift these recursion
relations to AdS. 

Our final answer for the $n$-point correlator $T_n$
is written schematically in the form:
\be
T_n = \sum_{\pi,\vect{e^{m'}}} \int_{{\cal H}} \left[ -i T^{*,\text{left}}_{m+1}(w) T^{*,\text{right}}_{n - m + 1}(w) +  \tilde{\cal B} \right] {d w \over w},
\ee
where the sum runs over various partitions of the operators into a ``left'' and a ``right'' set, and over the various
possible polarizations of an auxiliary ``internal particle'',  and the integral runs over a specified contour ${\cal H}$. In odd boundary dimensions, as we show in \cite{Raju:2012zs}, the integral over $w$ can be performed just by extracting residues at easily identifiable poles.  The $n$-point correlator factorizes into sums of products of ``transition amplitudes'' which are correlation functions taken between specified states as
discussed in \cite{Raju:2010by,Raju:2011ed}; this is why we place a $*$ in the superscript on the right hand side. $\tilde{\cal B}$ is a ``boundary term'' that is fixed by the Ward identities.

The second question we address in this paper is: can the flat space graviton amplitude be recovered from the boundary stress tensor correlator? This question was addressed in the early days of AdS/CFT --- albeit in a somewhat formal manner (see 
\cite{Polchinski:1999ry,Susskind:1998vk,Gary:2009ae,Gary:2009mi,Giddings:1999qu} and references there) --- and more recently
 in Mellin space \cite{Penedones:2010ue,Fitzpatrick:2011hu} where several explicit results were obtained. However, extending an observation first made for three-point functions in  \cite{Maldacena:2011nz}, we show that the flat-space limit is particularly elegant in momentum space: the flat-space graviton amplitude in $d+1$ dimensions appears as the coefficient of a specific singularity in the stress tensor correlator. 

In section \ref{secflatrecurs}, we prove that, for a scattering process at $l$-loops in pure gravity, the
flat space amplitude $M$ with polarization tensors $\vect{e^m}$ and on-shell momenta $\vect{\tilde{k}^m} = \{\vect{k^m}, i \norm{k^m} \}$ is related to the
the stress-tensor correlator $T$ by 
\be
M(\vect{e^1}, \vect{\tilde{k}^1}, \ldots \vect{e^n}, \vect{\tilde{k}^n}) = \lim_{E_T \rightarrow  0} {(E_T)^{\alpha_{\text{gr}}^l(n)}  \over \left(\prod \norm{k^m} \right)^{d - 1 \over 2} \Gamma(\alpha_{\text{gr}}^l) } T(\vect{e^1}, \vect{k^1}, \ldots \vect{e^n}, \vect{k^n}),
\ee
with $\alpha_{\text{gr}}^l(n) = {({n \over 2} - 1 + l)(d - 1) + 1},$ and $E_T = \sum \norm{k^m}$. 

In exactly the same way, the flat-space gluon scattering amplitude (with external polarization vectors $\vect{\ep^m}$)
is related to the current-correlators,
\be
M(\vect{\ep^1}, \vect{\tilde{k}^1}, \ldots \vect{\ep^n}, \vect{\tilde{k}^n}) = \lim_{E_T \rightarrow  0} {(E_T)^{\alpha_{\text{gl}}^l(n)}  \over \left(\prod \norm{k^m} \right)^{d - 3 \over 2} \Gamma(\alpha_{\text{gl}}^l) } T(\vect{\ep^1}, \vect{k^1}, \ldots \vect{\ep^n}, \vect{k^n}),
\ee
 although the singularity
now appears with an exponent
\be
\alpha_{\text{gl}}^l(n) = {({n \over 2} - 1 + l)(d - 3) + 1}.
\ee 
At higher than tree-level (i.e for $l > 0$), the relation above must be understood in dimensional regularization since both sides are UV-divergent.

The idea behind this limit is quite simple. Given a $d$-dimensional boundary momentum $\vect{k}$, we can append its norm to the vector and create a new $d+1$ dimensional {\em massless} momentum vector $\vect{\tilde{k}}$. The $d+1$ dimensional flat-space amplitude depends on these massless-momenta but involves momentum conservation in all $d+1$ dimensions. The boundary correlator conserves momentum only in $d$-dimensions. However, when we tune the boundary
momenta so that momentum in the ``radial'' direction is also conserved, then we get a singularity in the correlator with a coefficient that is precisely the flat space scattering amplitude!

Our flat space limit is valid more generally than our recursion relations. For one, it applies even at loop level in the bulk, although our recursion
relations are valid only at tree level. Second, it is straightforward to generalize it to the case of higher derivative interactions in the bulk as we describe below. So we hope that it will be of relevance more broadly beyond serving as a check on our answers for correlators. The flat space limit is also logically independent of the recursion relations, so the reader who is interest only in this aspect of the paper should skip to section \ref{secflatspace}.

A brief overview of this paper is as follows. In section \ref{secflatrecurs},
we present new recursion relations for graviton and gluon scattering in flat space. In section \ref{secrecursads}, we generalize these 
recursion relations to tree-level correlation functions of the stress-tensor or of conserved currents. In section \ref{secflatspace}, we prove
the flat space limit described above. In section \ref{secflatspaceofrecurs}, we
bring these two streams together and show that our recursion relations
automatically have the correct flat space limit. In the Appendix, we briefly discuss some of the problems associated with generalizing the usual BCFW recursion relations to $d=3$.

\section{Setting}
In this paper, we will consider correlation functions of the stress-tensor, and of conserved currents, in momentum space:
\be
 \langle T^{i_1 j_1}(\vect{k^1}) \ldots T^{i_n j_n}(\vect{k^n}) \rangle \equiv  \int \langle {\cal T}\Big\{T^{i_1 j_1}(\vect{x^1}) \ldots T^{i_n j_n}(\vect{x^n}) \Big\} \rangle e^{i \sum_{m=1}^n\vect{k^m} \cdot \vect{x^m}} d^d x_m, 
\ee
where ${\cal T}$ is the time-ordering symbol. 

It is convenient to think of this object as a functional of ``polarization'' tensors.
\be
\label{tdefstress}
T(\vect{e^1}, \vect{k^1}, \ldots \vect{e^n}, \vect{k^n}) = e^1_{i_1 j_1} \ldots e^n_{i_n j_n} \langle T^{i_1 j_1}(\vect{k^1}) \ldots T^{i_n j_n}(\vect{k^n}) \rangle,
\ee
where the $i_m, j_m$ run over the boundary directions.
For current correlators we can consider 
\be
\label{tdefcurr}
T(\vect{\ep^1}, \vect{k^1}, \ldots \vect{\ep^n}, \vect{k^n}) = \ep^1_{i_1} \ldots \ep^n_{i_n} \langle j^{i_1}(\vect{k^1}) \ldots j^{i_n}(\vect{k^n}) \rangle,
\ee
where we have suppressed the color indices carried by the currents, which will have no relevance in our analysis. 

However, note that in \eqref{tdefcurr}, if we have $\vect{\ep^n} = \vect{k^n}$, then the right hand side can be evaluated using Ward identities, which relate it to a lower point function. Similarly, in \eqref{tdefstress} if either (a)  $e^n_{i j} = v_{i} k_{j}$, for some $v_i$ or (b) $e^n_{i j} = -e^n_{j i}$ or (c) $e^n_{i j} = \eta_{i j}$ then the right hand side is determined in terms of various Ward identities \cite{Osborn:1993cr}.

This means that we only need to consider transverse-polarization vectors in \eqref{tdefcurr} and only symmetric, traceless, transverse polarization matrices in \eqref{tdefstress}. In $d$-dimensions, this allows $d-1$ polarization-vectors for currents and ${d (d - 1) \over 2} - 1$ polarization tensors for stress tensors.

If we are given the bulk action, we can compute these correlators directly using Witten diagrams \cite{Gubser:1998bc,Witten:1998qj}. However,  the Hilbert action of general relativity, when expanded in small fluctuations of the metric-tensor, leads to an infinite set of interaction vertices of increasing complexity.  So, in dealing with gravitational theories, it is necessary to find more efficient ways of computing these correlators, as we do below.

\paragraph{Notation:} In this paper, we use bold-face for vectors but not their components. The particle-number index
 on momenta or polarization vectors is usually placed in the superscript 
and we usually use $m,n$ etc for this index. We use  $i,j$ etc. for boundary spacetime indices and $\mu, \nu$ etc. for bulk spacetime indices

\section{New Flat Space Recursion Relations \label{secflatrecurs}}
In this section we start by describing some new recursion relations in flat space. These will help establish notation and serve as a warm-up for the new recursion relations in AdS. We first describe these recursion relations for gauge-boson amplitudes, and then for graviton amplitudes.

\subsection{Recursion in Yang-Mills}
Consider an amplitude in Yang-Mills theory --- $M(\vect{k^1}, \vect{\ep^1} \ldots \vect{k^n}, \vect{\ep^n})$ ---  where the external gluons have momenta $\vect{k^m}$ and polarizations $\vect{\ep^m}$. In order to apply the recursion relations, we will need the further constraints that some set of $m$ of these vectors is linearly dependent. Without loss of generality, we can take these to be first $m$-insertions.
\be
\label{constraint}
\sum_{p=1}^m \alpha_{p} \vect{\ep^{p}} = 0,
\ee
where the $\alpha$ are some coefficients. Now, polarization vectors can be shifted by a multiple of the momentum. In 4-dimensional theories, for any 4-point and higher amplitude, we can always use this freedom to find a set of polarizations that satisfy \eqref{constraint}. In higher dimensions, we can build up an amplitude with more general polarization vectors, by using linear combinations of polarizations that satisfy \eqref{constraint} as explained in section 4.4 of \cite{Raju:2011ed}. 

Now, consider deforming the amplitude through the extension
\begin{equation}
\label{risagerextension}
\vect{k^p} \rightarrow \vect{k^p}(w) \equiv \vect{k^p} + \alpha_p \vect{\ep^p} w, \quad p \leq m
\end{equation}
for {\em each} of the first $m$-insertions. Note that there is no sum over $p$ in the second term above. The condition \eqref{constraint} ensures that momentum is conserved under this deformation. This is similar to the extension described by Risager \cite{Risager:2005vk}.

The tree-amplitude is a rational function of $w$, and it is quite easy to see that it dies off at large $w$. To see this, were merely need to apply the Ward identities. For large $w$,
\be
\label{largewYM}
\begin{split}
&M(\vect{k^1}(w), \vect{\ep^1} \ldots \vect{k^m}(w), \vect{\ep^m}(w), \ldots  \vect{k^n}, \vect{\ep^n}) \\&= \ep^{1}_{\mu_1}  \ldots \ep^{m}_{\mu_m} \ldots  \ep^{n}_{\mu_n} M_F^{\mu_1 \ldots \mu_m \ldots \mu_n}(\vect{k^1}(w), \ldots \vect{k^m}(w), \ldots  \vect{k^n})  \\
&= {{\cal N} \over w^m} \left({k^{1}_{\mu_1}(w)  - k^{1}_{\mu_1}(0)}\right)\ldots  \left(k^{m}_{\mu_m}(w) - k^{m}_{\mu_m}(0)\right) \ldots  \ep^n_{\mu_n}  \\
&\times M_F^{\mu_1 \ldots \mu_m \ldots \mu_n}(\vect{k^1}(w), \ldots \vect{k^m}(w), \ldots  \vect{k^n}),
\end{split}
\ee
where $M_F^{\mu_1, \ldots \mu_n}$ comes from summing all Feynman diagrams that contribute to the amplitude and we have defined ${\cal N} = \left(\prod \alpha_i\right)^{-1}$.   However, the Ward identities tell us that whenever we contract a momentum with $M_F$ we get zero. Moreover, $M_F$ itself can, at worst, scale like $w$ under the extension \eqref{risagerextension}. So the expression in \eqref{largewYM} vanishes at large $w$. 

As usual the poles of the amplitude under \eqref{risagerextension} occur whenever an intermediate propagator goes on shell. The residue at such a pole is a product of the left and the right amplitudes. We can reconstruct the full amplitude from a knowledge of these residues. 

This leads to the following recursion relations\footnote{If not all the momenta are shifted i.e. if $m < n$,  then, in what follows, we interpret $\vect{k}_o(w) = \vect{k}_o$ for $o > m$.}
\begin{equation}
\label{ymrecurs}
\begin{split}
&M(\vect{\ep^1}, \vect{k^1}(w), \ldots \vect{\ep^n}, \vect{k^n}(w)) = \sum_{\{\pi\},h, \pm}  {-i {\cal M}^2 \over (\sum_{o=1}^{m_l} \vect{k^{\pi_o}}(w))^2 }  {w - w^{\mp} \over w^{\pm} - w^{\mp}}, \\
&{\cal M}^2 \equiv  {M(\vect{\ep^{\pi_1}}, \vect{k^{\pi_1}}(w^{\pm}),  \ldots \vect{\ep^{q'}_h}, \vect{k^{q'}}) M(\vect{\ep^{q'}_{-h}}, -\vect{k^{q'}},\ldots \vect{\ep^n}, \vect{k^{\pi_n}}(w^{\pm})) }.
\end{split}
\end{equation}
We need to explain several parts of this expression. 
\begin{enumerate}
\item
First, let us examine the sum over $\pi$.  This sum is over all ways of partitioning the external momenta into two sets -- $\{\pi_1, \ldots \pi_{m_l}\}, \{\pi_{m_{l+1}}, \ldots \pi_n\}$. We will call these sets,  ``left'' and ``right'' below; they  have the property that each set contains {\em at least one} of the first $m$-momenta. 
\item
Each such partition is in one-to-one correspondence with poles in the integrand of the amplitude. To describe this relation, we define
\be
\vect{k^{q'}} = \sum_{o=1}^{m_l} \Big[ \vect{k^{\pi_o}} + \theta(m-\pi_o) \alpha_{\pi_o} \vect{\ep^{\pi_0}} w^{\pm} \Big].
\ee
This is just the sum of all the extended momenta in the left partition, where the $\theta$ function ensures that only the first $m$-momenta are extended.
The complex numbers $w^{\pm}$ are now defined by solving the equation:
\be
(\vect{k^{q'}})^2 = 0.
\ee
There are two solutions because this is a quadratic equation in $w$. This is what leads to the funny-looking factor of ${w^{\mp} \over w^{\pm} - w^{\mp}}$.
\item
The sum over intermediate helicities $h$ leads to the insertion of 
any complete set of polarization vectors for the intermediate particle i.e. while contracting with on-shell amplitudes on the left and the right, the following replacement should be allowed:
\be
\sum_{h} \ep^{q'}_{h,\mu} \ep^{q'}_{-h,\nu} \rightarrow \eta_{\mu \nu}.
\ee
\end{enumerate}

\subsection{Recursion in Gravity \label{secrecursgravity}}
We now turn to a description of how these new recursion relations can
be implemented for theories of gravity.  There are two  differences from the case of Yang-Mills explained above: the first has to do with the conditions on polarization tensors, and the second has to do with the large $w$ behaviour.

\begin{figure}
\label{gravitonslargew}
\begin{center}
\includegraphics[height=5cm]{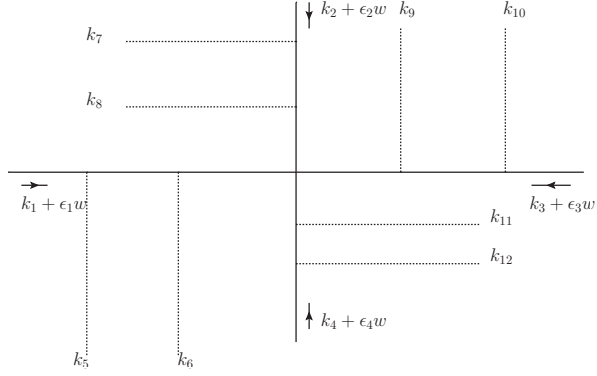}
\end{center}
\caption{Gravity Feynman diagram, with 4 momenta of $O(w)$, that scales like $w^{n-2}$}
\end{figure}

For the recursion relations to apply, we require the following condition. Some $m$ of the polarization tensors should be writable as:
\be
e^q_{\mu \nu} = \ep^q_{(\mu} v^q_{\nu)},
\ee
where the $\vect{\ep^q}$ satisfy:
\be
\vect{\ep^q} \cdot \vect{\ep^q} = 0, \quad \vect{\ep^q} \cdot \vect{k^q} = 0,
\ee
and are linearly dependent as in \eqref{constraint}.
For $\vect{e}$ to be a valid polarization tensor, we must have
\be
\vect{\ep^q} \cdot \vect{v^q} = 0, \quad \vect{v^q} \cdot \vect{k^q} = 0.
\ee
Second, for the correlator to die off at large $w$, the number of particles extended according to \eqref{risagerextension}, say $m$, must have the property that 
\be
\label{stressmcondition}
2 m > n+2.
\ee
This is because, a gravity Feynman-diagram with $m$ momenta scaling like $w$ can naively scale as fast as $w^{n+2-m}$ as shown in figure \ref{gravitonslargew} which shows an example in the case where $m=4$. In this diagram all the four solid lines have propagators that are $O(w)$ (because the $\vect{\ep^q}$ are null vectors) but interaction vertices that are $O(w^2)$.

With these caveats, we can repeat the argument above to obtain the following recursion relations for graviton amplitudes in flat space. 
\begin{equation}
\label{gravityrecurs}
\begin{split}
&M(\vect{e^1}, \vect{k^1}(w), \ldots \vect{e^n}, \vect{k^n}(w)) = \sum_{\{\pi\},h, \pm}  {-i {\cal M}^2 \over p^2 + (\sum_{o=1}^{m_l} \vect{k^{\pi_o}}(w))^2}   {w - w^{\mp}(p) \over w^{\pm}(p) - w^{\mp}(p)}, \\
&{\cal M}^2 \equiv  {M(\vect{e^{\pi_1}}, \vect{k^{\pi_1}}(p),  \ldots \vect{e^{q'}_h}, \vect{k^{q'}}) M(\vect{e^{q'}_{-h}}, -\vect{k^{q'}},\ldots \vect{e^n}, \vect{k^{\pi_n}}(p)) }.
\end{split}
\end{equation}
The notation used here is exactly the same as the notation for \eqref{ymrecurs}. The intermediate polarization tensors $\vect{e^{q'}_{\pm h}}$ runs over any complete set of graviton polarizations.

\section{Recursion Relations in AdS \label{secrecursads}}
We now turn to a description of how these recursion relations can be generalized to AdS. We first need to discuss the behaviour of current and stress-tensor correlators under the extension \eqref{risagerextension}.

\subsection{Large $w$ Behaviour for Current Correlators}
In this subsection we show that under \eqref{risagerextension}, a current correlator vanishes at large $w$.

The correlator we are interested in is
\begin{equation}
\label{currentcor}
T = \ep^{1}_{i_1} \ldots  \ep^{m}_{i_m} \langle j^{i_1}(\vect{k^1}(w)) j^{i_m}(\vect{k^m}(w)) O(\vect{k^{m+1}}) \ldots O(\vect{k^n})\rangle.
\end{equation}
Here, the operators that carry the momenta that we are deforming are denoted by $j$, and we have denoted all the  ``other'' operators that might exist in the correlator by $O$. 

We can now substitute for the polarization vectors in terms of the extended and un-extended momenta as in the Yang-Mills analysis above:
\begin{equation}
\label{overallw}
T 
= {{\cal N} \over w^m} \left({k^{1}_{i_1}(w)  - k^{1}_{i_1}(0)}\right) \left(k^{m}_{i_m}(w) - k^{m}_{i_m}(0)\right)  \langle j^{i_1}(\vect{k^1}(w)) \ldots j^{i_m}(\vect{k^m}(w)) O(\vect{k^{m+1}}) \ldots  O(\vect{k^n}) \rangle.
\end{equation}

We would like to use the fact, that for large $w$, the polarization vectors are approximately proportional to the momentum, to simplify the correlator. However, we need to be more careful about this argument since the Ward identities for correlation functions can give contact terms on the right hand side.

In fact the application of the Ward identities gives the two sorts of terms shown below. 
\begin{align}
\label{wardidentityapplication}
&k^{1}_{i_1}(w)  k^{2}_{i_2}(w) \ldots k^{m}_{i_m}(w)\langle j^{i_1}(\vect{k^1}(w))  j^{i_2}(\vect{k^2}(w)) \ldots j^{i_m}(\vect{k^m}(w))  O(\vect{k^{m+1}}) \ldots O(\vect{k^n})\rangle \\ \label{k1ink2} &=  k_{i_2}^2(w) \ldots  k_{i_m}^m(w) \Big(\langle j^{i_2}(\vect{k^2}(w) + \vect{k^1}(w)) \ldots  j^{i_m}(\vect{k^m}(w))  O(\vect{k^{m+1}}) \ldots O(\vect{k^n}) \rangle  \\ \label{k1inO} & + \langle j^{i_2}(\vect{k^2}(w))  \ldots  j^{i_m}(\vect{k^m}(w))  O(\vect{k^{m+1}} + \vect{k^1}(w)) \ldots  O(\vect{k^n}) \rangle + \ldots  \Big),
 \end{align}
The first kind of terms are those where the $\vect{k_1(w)}$ moves into one of the other $j$ operators, and the second kind are where the $\vect{k_1(w)}$ moves into one of the $O$ operators. Note that in \eqref{k1ink2} we cannot, any more, use the Ward identity to contract with $\vect{k_2(w)}$, whereas we can do this in \eqref{k1inO}. Proceeding in this way, we come to a situation
where we have several terms, each of which has the following form: it has $t$ polarization vectors that scale with $w$ left on the outside, $t$ of the $j$ operators have momenta that scale with $w$, and $m-2 t$ of the $O$-operators have picked up momenta that scale with $w$. 

In any such term, the correlator itself, barring the polarization vectors, has a total of $m-t$ momenta scaling like $w$. It is easy to persuade oneself that in Yang-Mills theory with no higher derivative terms this correlator cannot scale any faster than than $w$.  After multiplying with the polarization vectors, we see that the expression in \eqref{wardidentityapplication} can at most contain terms that scale as $w^{t+1}$. Hence, the correlator in \eqref{overallw} reduces to terms that die off like $w^{t+1-m}$ at large $w$.

\subsection{Large $w$ Behaviour for Stress-Tensor Correlators}
We now turn to the case of stress-tensor correlators. We will find below
that, in fact, stress tensor correlation functions do not die off at large $w$. However, the behaviour at large $w$ is entirely determined by the Ward identities.

For stress-tensor correlators, we make the substitution
\be
\label{largewstresssubst}
e^{q}_{i j} = {1 \over 2} {k^{q}_{i}(w) - k^{q}_{i}(0) \over \alpha_q w} v^{q}_{j} + i \leftrightarrow j
\ee
for the first $m$ polarization tensors in \eqref{tdefstress}. 

However, unlike the case of current correlators, we cannot use this substitution to argue that there are no terms at large $w$. The Ward identities for the four-point function can be worked out in a straightforward manner following \cite{Osborn:1993cr} (although their exact
form also depends on the precise definition of the correlator.) However, all that is important to us is that we find terms of the sort:
\be
T(\vect{k^1}, \vect{\ep^1} \otimes  \vect{k^1}, \ldots \vect{k^n}, \vect{e^n}) =  \sum_{q} \vect{k^q} \cdot \vect{\ep^1}  T(\vect{k^2}, \vect{e^2}, \ldots  \vect{k^q}, \vect{e^q}, \ldots \vect{k^n},\vect{e^n}) + \ldots
\ee
However  $\vect{\ep^1} \cdot \vect{k^q}$ could grow with $w$, since $\vect{k^q}$ grows with $w$, if $q < m$ under the deformation \eqref{risagerextension}. However, while this term does not vanish at large $w$, its behaviour is completely fixed by the Ward identities.

Let us state this a little more precisely. We can write 
\be
\begin{split}
I(w) =  {{\cal N} \over w^m} &\left[\prod_q (k^q_{ i_q}(w) - k^q_{ i_q}(0)) v^{q}_{j_q} - (-1)^m \prod_q k^q_{ i_q}(0) v^{q}_{j_q} \right] \\ &\times \langle T^{i_1 j_1}(\vect{k^1}) \ldots T^{i_m j_m}(\vect{k^m}) O(\vect{k^{m+1}}) \ldots O(\vect{k^n}) \rangle.
\end{split}
\ee
$I(w)$ is {\em completely determined} by the Ward identities and our knowledge of lower-point functions.  

So, if we substitute \eqref{largewstresssubst} into the correlator, there is exactly one term that is not determined in this way. This is the term
\[
{1 \over w^m} \left[\prod_q k^q_{(i_q}(0) v^{q}_{j_q)} \right] \langle T^{i_1 j_1}(\vect{k^1}) \ldots T^{i_m j_m}(\vect{k^m}) \ldots O(\vect{k^{m+1}}) \ldots O(\vect{k^n}) \rangle.
\]
This term vanishes at large $w$ provided that the ``bare correlator'' does
not grow any faster than $w^m$. Now, as we pointed out above in the analysis for graviton scattering amplitudes, the bare correlator with $m$-momenta extended may grow as fast as $w^{n+2-m}$. So,  provided
\be
2 m > n + 2,
\ee
the large-$w$ behaviour of the stress-tensor correlator under \eqref{risagerextension} is completely determined.

\subsection{Recursion Relations for Currents}
Repeating the arguments of \cite{Raju:2010by, Raju:2011ed}, we find that
we now have the following information about correlation functions of currents that are dual to tree-level Witten diagrams of Yang-Mills theory in the bulk:
\begin{enumerate}
\item
Under the extension \eqref{risagerextension}, these correlators can be 
written as integrals of a rational function of $w$. The integration
variables are $n-3$ parameters,\footnote{The counting of $n-3$ comes from the diagrams that involve three-point interactions joined together with bulk to bulk propagators. However, 
even if have four or higher point interactions, each Witten diagram can always
be written in this form. See section 6 of \cite{Raju:2012zs} in the neighbourhood of equation 6.28.}  each of which comes from an integral over $p$ in the bulk-bulk propagators. (We are adopting the same notation as \cite{Raju:2010by,Raju:2011ed} but the bulk-bulk propagator is also shown explicitly in \eqref{gravitypropagator}.)
\item
The only poles in this integrand occur when the denominator of one of the
bulk-bulk propagators vanishes. At this point, the residue of the pole
is the product of the integrands of two smaller ``transition amplitudes'' i.e. the quantities obtained by replacing one bulk to boundary propagator in a Witten diagram by a normalizable mode.
\item
At large $w$, the behaviour of the integral is controlled by the discussion above. 
\end{enumerate}
This leads to the following recursion relations.
\begin{equation}
\label{currecurs}
\begin{split}
&T(\vect{\ep^1}, \vect{k^1}(w), \ldots \vect{\ep^n}, \vect{k^n}(w)) = \sum_{\{\pi\},\vect{\ep^{q'}}. \pm}\int  \left[{-i {\cal T}^2 \over p^2 + (\sum_{o=1}^{m_l} \vect{k^{\pi_o}}(w))^2} {w - w^{\mp}(p) \over w^{\pm}(p) - w^{\mp}(p)} + {\cal B} \right] {d p^2 \over 2}, \\
&{\cal T}^2 \equiv  {T^*(\vect{\ep}^{\pi_1}, \vect{k^{\pi_1}}(p),  \ldots \vect{\ep^{q'}}, \vect{k^{q'}}) T^*(\vect{\ep^{q'}}, -\vect{k^{q'}},\ldots \vect{\ep^{\pi_n}}, \vect{k^{\pi_n}}(p)) }.
\end{split}
\end{equation}
Although we have written the expression for arbitrary $w$, we will often
only be interested in the value of the correlator at $w = 0$. The notation above is the same as the notation used in \eqref{ymrecurs}.  The $T^*$ is an amplitude for all the ``left'' insertions to go into an ``intermediate state'' with momentum $\vect{k_q'}$ defined above. We have placed a $^*$ in the superscript of $T$ to emphasize that this is a {\em transition amplitude.} It is computed by using the usual bulk-boundary propagators for all particles indexed by $\pi_1, \ldots \pi_{m_l}$, but by using a {\em normalizable mode} for the particle with momenta $\vect{k_q'}$. These quantities were first
described in \cite{Balasubramanian:1999ri} and are also discussed in 
detail in \cite{Raju:2011ed}. 

Finally, ${\cal B}$ is a boundary term that is required to fix the behaviour of the integrand at large $p$ and large $w$.  The fact that the term with ${\cal T}^2$ already correctly reproduces the poles of the integrand at finite $w$ tells us that ${\cal B}$ must be of the form
\be
{\cal B} = \sum_{m=0} a_m(p) w^m,
\ee
where the $a_m(p)$ are some rational functions. 
If the term involving ${\cal T}^2$ grows at large $p$, we must use ${\cal B}$ to cancel this growth since we know that the $p$-integrals in the bulk to bulk propagators that we started with are all convergent. Second, since the behaviour of the integral at large $w$ is fixed by the discussion on Ward identities above, we also know the integrals of the functions $a_m(p)$. This fixes ${\cal B}$ up to irrelevant terms that integrate to $0$.  We will see in \cite{Raju:2012zs} that, at the level of four point functions in AdS$_4$/CFT$_3$, we never need to evaluate ${\cal B}$ explicitly.

\subsection{Recursion Relations for Stress Tensors}
We now turn to the case of stress-tensor correlators. These correlators
are labeled by a momentum, and transverse-traceless polarization tensors just like graviton amplitudes. For our recursion relations to apply we require the conditions that were enumerated in \hbox{section \ref{secrecursgravity}}. With these constraints on the polarization tensors, we find that
\begin{equation}
\label{stressrecurs}
\begin{split}
&T(\vect{e^1}, \vect{k^1}(w), \ldots \vect{e^n}, \vect{k^n}(w)) = \sum_{\{\pi\},\vect{e^{q'}}, \pm} \int  \left[{-i {\cal T}^2 \over p^2 + (\sum_{o=1}^{m_l} \vect{k^{\pi_o}}(w))^2} {w - w^{\mp}(p) \over w^{\pm}(p) - w^{\mp}(p)} + {\cal B} \right] {d p^2 \over 2}, \\
&{\cal T}^2 \equiv  {T^*(\vect{e^{\pi_1}}, \vect{k^{\pi_1}}(p),  \ldots \vect{e^{q'}}, \vect{k^{q'}}) T^*(\vect{e^{q'}}, -\vect{k^{q'}},\ldots \vect{e^{\pi_n}}, \vect{k^{\pi_n}}(p)) }.
\end{split}
\end{equation}
The notation is the same as that used above. 

\subsection{Another form of the relations}
Let us specialize the recursion relations to $w=0$. Then we can rewrite 
\eqref{stressrecurs} following \cite{ArkaniHamed:2009si} 
\be
\begin{split}
&\sum_{\pm} \int_0^{\infty}  {i {\cal T}^2  \over p^2 + (\sum_{o=1}^{m_l} \vect{k^{\pi_o}})^2}  {d p^2 \over 2} {w^{\mp}(p) \over w^{\pm}(p) - w^{\mp}(p)} \\ &= \int_0^{\infty}  {d p^2 \over 2} \int_{\cal H} {d w \over w}  \Bigg[-i {\cal T}^2  \delta\Big(\big(\sum_{o=1}^{m_l} \vect{k^{\pi_o}}(w)\big)^2 + p^2\Big) \Bigg],
\end{split}
\ee
where ${\cal H}$ is the set of points on the $w$ plane that satisfy
\be
{\rm Im}\Big[(\sum_{o=1}^{m_l} \vect{k^{\pi_o}}(w))^2\Big]= 0,\quad \text{and} \quad {\rm Re}\Big[(\sum_{o=1}^{m_l} \vect{k^{\pi_o}}(w))^2\Big] < 0 \quad \text{for}\quad w \in {\cal H}.
\ee
This is the intersection of the union of the two curves that solve the quadratic equation with the region that satisfies the inequality.

We can check this relation by just doing the integral over the $\delta$ function.  If we
write $Q(w) = \left(\sum_{o=1}^{m_l} \vect{k^{\pi_o}(w)}\right)^2 + p^2 = A (w - w^+) (w - w^-)$, then\footnote{The relative sign we get between the contribution from $w^{+}$ and $w^{-}$ is sensitive to the direction along which we integrate along the contour ${\cal H}$.}
\be
\int_{\cal H} {d w \over w} {\cal T}^2(w) \delta \Big(Q(w) \Big)  = \sum_{\pm} {\cal T}^2(w^{\pm}) {\delta(w - w^{\pm}) \over A w^{\pm} (w^{\pm} - w^{\mp})} = {1 \over Q(0)} \sum_{\pm} {\cal T}^2(w^{\pm}) {w^{\mp} \over w^{\pm} - w^{\mp}}.
\ee

We can now interchange the order of integration, do the integral over $p$, and rewrite the relations with only an integral over $w$. 
\be
\label{stressrecursw}
T(\vect{e^1}, \vect{k^1}, \ldots \vect{e^n}, \vect{k^n}) = \sum_{\{\pi\},\vect{e^{m'}}}\int_{\cal H}  \left[{-i {\cal T}^2 \over w} + \tilde{{\cal B}} \right] d w,
\ee
where $\tilde{{\cal B}}$ just comes from rewriting ${\cal B}$ as a function of $w$ and multiplying with the Jacobian factor for the change of variables from $p$ to $w$.

Although this expression is somewhat neater than \eqref{stressrecursw} 
it has the disadvantage that the contour ${\cal H}$ can be somewhat complicated. We should remind the reader that the momenta on the left hand side are not deformed, and $w$ on the right hand side is a dummy variable that is integrated over.\section{A New Flat Space Limit \label{secflatspace}}
In this section, we would like to describe a new flat space limit
of AdS correlators, which relates the $d$-dimensional
correlator of stress-tensors, or of currents,  computed using Witten diagrams, and the $d+1$-dimensional flat-space amplitude of gravitons or gluons.

Before we describe this limit, it is useful to review the analytic 
structure of $d$-dimensional correlators in momentum space. 
Now, the bulk to boundary propagators in AdS are given by the following expressions:
\begin{equation}
\label{solgravfree}
h_{i}^{j}(\vect{e}, \vect{k}, \vect{x},z) =  \sqrt{2 \over \pi}  e_{i}^{j} (\norm{k} z)^{{d \over 2}} e^{i \vect{k} \cdot \vect{x}}  K_{d \over 2}(|\vect{k}| z),
\end{equation}
where
\be
h_{0\mu} = 0, \quad k_i e^{i j} = 0, \quad e^{i}_{i} = 0.
\ee
It is important to note that in \eqref{solgravfree}, we have raised one index on $h$. If both indices were lowered, we
would have an additional factor of $z^{-2}$ on the right hand side. Here $\norm{k}$ is chosen to be positive if $\vect{k}$ is spacelike
and it is chosen to have negative imaginary part if $\vect{k}$ is timelike. The physical computation in AdS requires these signs.

 We also need the
bulk-bulk propagator that, for gravity in axial gauge, is given by \cite{Raju:2011ed}:
\be
\label{gravitypropagator}
G_{i l}^{j k} =
\int \left[{
e^{i \vect{k} \cdot (\vect{x} - \vect{x'})} 
( z z' )^{d \over 2} J_{d \over  2}(p z) J_{d \over 2} (p z') \over  
\left(\vect{k}^2 + p^2 - i \epsilon\right)} \right.   {1 \over 2} \left.\left({\cal T}_{i}^{k} {\cal T}^{j}_{l} + {\cal T}_{i l} {\cal T}^{j k} - 
{2 {\cal T}_{i}^{j} {\cal T}^{k}_{l}\over d-1} \right)\right] {-i d^d \vect{k} d p^2 \over 2 (2 \pi)^d},  
\end{equation}
where ${\cal T}_{i}^{j} = \delta_{i}^{j} + k_{i} k^{j}/p^2$, and the $i,j$ indices are raised and lowered using the flat-space $d$-dimensional metric.

A typical Witten diagram such as the one shown in figure \ref{figadsexchange} or figure
\ref{figadsoneloop} involves several radial integrals and integrals over
the radial momenta $p$ in the bulk-bulk propagators. 
{\em After}
we have done all the radial integrals, we are left with various integrals over $p$. At this stage, we are free to analytically
continue and  flip the sign of $\norm{k}$. This leads to the function
\[
T(\vect{k^1}, \norm{k^1}, \vect{e^1}, \ldots \vect{k^n}, \norm{k^n}, \vect{e^n}),
\]
which depends on the polarizations $\vect{e^m}$, the three-momenta $\vect{k^m}$ 
and their norms and where there is no constraint on the sign of $\norm{k^m}$, although we still demand that $\norm{k^m}^2 = \vect{k^m} \cdot \vect{k^m}$.

We can also consider forming the $d+1$-dimensional null momentum:
\be
\label{tildekdef}
\vect{\tilde{k}} = (\vect{k}, i \norm{k}).
\ee
The $d+1$ dimensional scattering amplitude naturally depends on 
these ``massless momenta'' and the external polarizations:
\[
M(\vect{\tilde{k}^1}, \vect{e^1}, \ldots \vect{\tilde{k}^n}, \vect{e^n})
\]
In what follows below we explore the relation between these two quantities --- $M$ and $T$.

 It will be convenient below for us to define the quantity
\be
\label{etdef}
E_T = \sum_{q=1}^n \norm{k^q},
\ee
which is the total ``radial momentum.'' The momentum conserving
delta functions in  the flat-space amplitude, of course, include a factor of $\delta(E_T)$.  We will show below that the coefficient of this $\delta$-function is just the residue of a pole at $E_T = 0$ in the CFT correlator.

In our explicit computations below, we take the bulk action to be either the pure Hilbert action for gravity or the Yang-Mills action. However, as we mention below it is not difficult to generalize our results to other
kinds of interactions.

\subsection{Flat Space Limit for Tree Amplitudes}
We\footnote{This subsection was worked out in collaboration with Juan Maldacena and Guilherme Pimentel. These results were presented in \cite{Maldacenastrings11}.} will now show that the $d$-dimensional stress-tensor correlator and the $(d+1)$-dimensional graviton tree-amplitude  are related through 
\be
\label{flatspacestress}
M(\vect{e^1}, \vect{\tilde{k}^1}, \ldots \vect{e^n}, \vect{\tilde{k}^n}) = \lim_{E_T \rightarrow  0} {(E_T)^{\alpha_{\text{gr}}^0(n)}  \over \left(\prod_{m=1}^n \norm{k^m} \right)^{d - 1 \over 2} \Gamma(\alpha_{\text{gr}}^0) } T(\vect{e^1}, \vect{k^1}, \ldots \vect{e^n}, \vect{k^n}),
\ee
with 
\be
\alpha_{\text{gr}}^0(n) = {({n \over 2} - 1)(d - 1) + 1},
\ee 
and $E_T$ defined in \eqref{etdef}.
A similar relation holds for current correlators: 
\be
\label{flatspacecurr}
M(\vect{\ep^1}, \vect{\tilde{k}^1}, \ldots \vect{\ep^n}, \vect{\tilde{k}^n}) = \lim_{E_T \rightarrow  0} {(E_T)^{\alpha_{\text{gl}}^0(n)} \over \left(\prod_{m=1}^n \norm{k^m}\right)^{d - 3 \over 2} \Gamma(\alpha_{\text{gl}}^0)} T(\vect{\ep^1}, \vect{k^1}, \ldots \vect{\ep^n}, \vect{k^n}),
\ee
with 
\be
\alpha_{\text{gl}}^0 = \left({n \over 2} - 1 \right)(d - 3) + 1.
\ee
In writing this relation, we are stripping off the overall momentum conserving delta functions on both sides. For the flat-space amplitude, momentum is conserved in all $d+1$ directions, whereas the correlator only conserves momentum in $d$-directions. The pole shown above occurs when the total $z$-momentum in the correlator also vanishes. 

We should mention that both sides of \eqref{flatspacestress} manifestly have the same behaviour under rescalings of the momenta. The $d$-dimensional
tree-level graviton scattering amplitude scales as $M \rightarrow \lambda^2 M$ if all the momenta are rescaled by $\vect{k^m} \rightarrow \lambda \vect{k^m}$. The stress tensor correlator, {\em without the leading $\delta$-function}, scales like $T \rightarrow \lambda^d T$ under this scaling. We see that the pre-factor equalizes the behaviour under scaling of both sides. Similarly, the $d$-dimensional gluon amplitude scales like $M \rightarrow \lambda^{4-n} M$, while the current correlator scales as $T \rightarrow \lambda^{d-n} T$; the pre-factor turns this scaling into that of the amplitude.

\subsubsection{Contact Interactions}
We start by discussing contact interactions and then go on to 
discuss interactions involving bulk propagators. The contribution of a 
contact interaction, such as the one shown in Fig. \ref{figcontact}, 
to the 
momentum space correlator can be written as the integral of a function of
$z$ and the momenta $\vect{k^i}$
\be
\label{adscontact}
T(\vect{k^i}) = \int C_a(z, \vect{k^i}) \sqrt{-g(z)} d z + \ldots 
\ee
where the $\ldots$ indicate other terms
that contribute to the correlator.

Now, in flat space, although we 
would usually choose to Fourier transform in the $z$ direction as well, we can write down a similar contribution leaving the $z$-integral as is: 
\be
\label{flatcontact}
M(\vect{\tilde{k}^m}) = \int {\cal C}_f(z, \vect{\tilde{k}^m}) d z + \ldots 
\ee
How are ${\cal C}_f$ and $C_a$ related?

In general, the answer to this is quite complicated, but to get the flat space limit, we are interested in what happens in the deep interior of AdS i.e. at large $z$. At large $z$, the relation between ${\cal C}_f$ and $C_a$ simplifies as we now show. 

Both $C_a$ and ${\cal C}_f$ are related to the contact vertex which is obtained by expanding the Hilbert action out to the appropriate power in a perturbation about AdS. However notice that
\be
\label{conformaltransricci}
R(g^{\rm ads}_{\mu \nu} + h_{\mu \nu}) = R\left({1 \over z^2} (\eta_{\mu \nu} + z^2 h_{\mu \nu})\right) = z^2 R(\eta_{\mu \nu} +  z^2 h_{\mu \nu}) - d (d+1).
\ee
\FIGURE[!h]{
\label{figcontact}
\includegraphics[height=5cm]{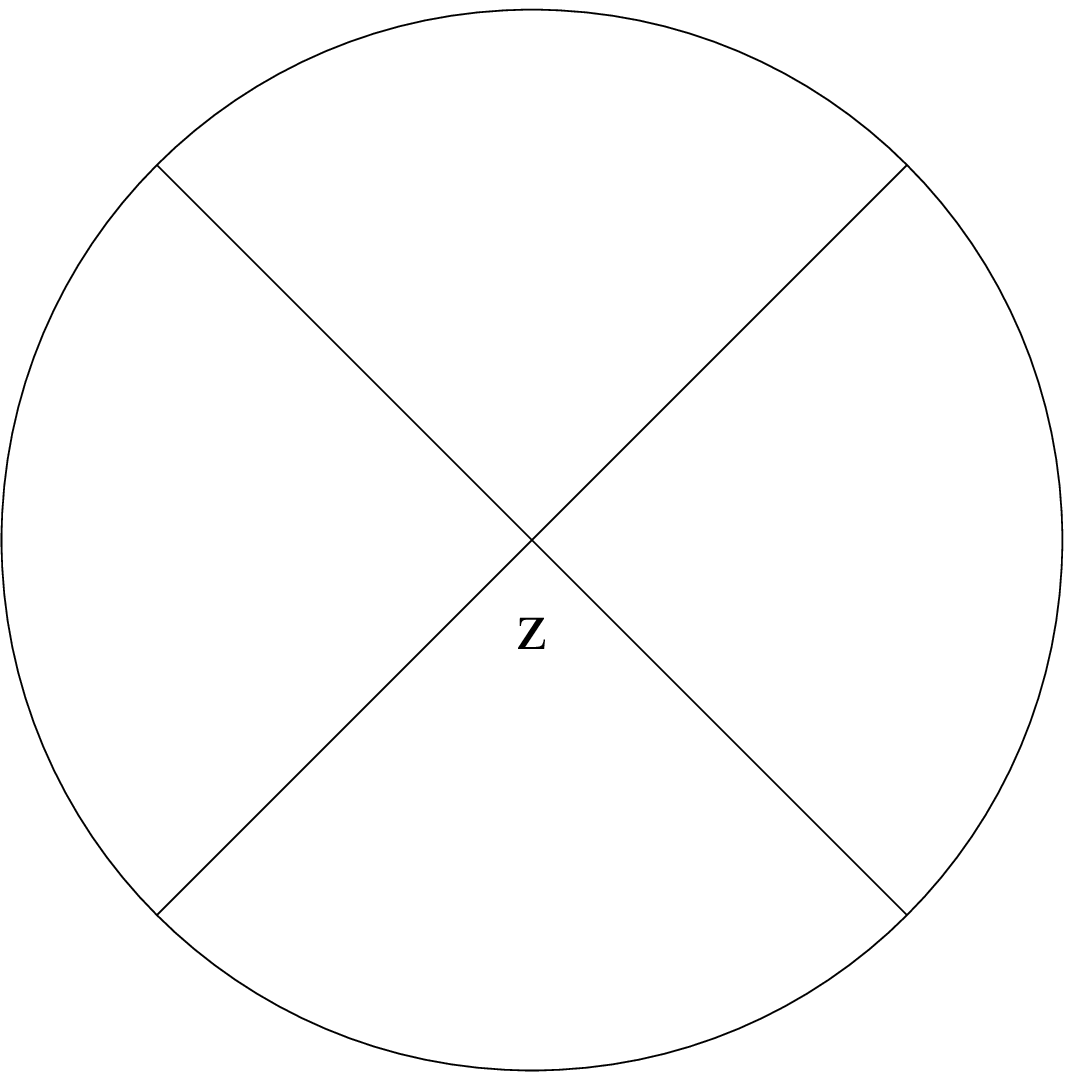}
\caption{Contact Interaction in AdS}
}

The wave functions described in \eqref{solgravfree} have the following behaviour at large $z$:
\be
z^2 h_{\mu \nu}(x, z) \underset{z \rightarrow \infty}{\longrightarrow} (\norm{k} z)^{d - 1 \over 2} e^{- \norm{k} z + i {k \cdot x}},
\ee

When we expand out the scalar curvature $R$ on the right hand side in \eqref{conformaltransricci} there are various $z$-derivatives that act on $h$. However, if we want to get the largest power of $z$ in a n-point contact 
interaction, then we must make sure that all the $z$-derivatives act only
on the exponential and not on the leading pre-factor. 
After we take into account the additional factor of $\sqrt{-g(z)}=z^{-d - 1}$, this leads to the result
\be
\label{contactvertexscaling}
\sqrt{-g(z)} C_a(z,\vect{k^m}) \underset{z \rightarrow \infty}{\longrightarrow} \left(\prod_{q=1}^n \norm{k^q}\right)^{d - 1 \over 2}  z^{({n \over 2} - 1)(d - 1) } {\cal C}_f(z, \vect{k^m}).
\ee

Now, there is another difference between \eqref{adscontact} and \eqref{flatcontact}, which involves the range of the $z$-integral. Both ${\cal C}_f$ and $C_a$ involve a leading exponential in $z$: $e^{-E_T z}$. Integrating 
this over all $z$ in \eqref{flatcontact} gives a $\delta$ function: $\delta(E_T)$. However, doing the integral from $0$ to $\infty$ in \eqref{adscontact} with 
the leading power of $z$ shown above leads to a pole at $E_T = 0$ and the relation \eqref{flatspacestress}.

\subsubsection{Exchange Interactions: Differential Equation Argument}
Now, a correlator  receives contributions not only from contact Witten diagram, but also from diagrams with bulk-bulk propagators. From the argument above, it is clear that the contact diagram yields the flat space result multiplied with the correct pole. We will now show that this happens for terms with propagators as well.

\FIGURE{
\includegraphics[width=4.5cm]{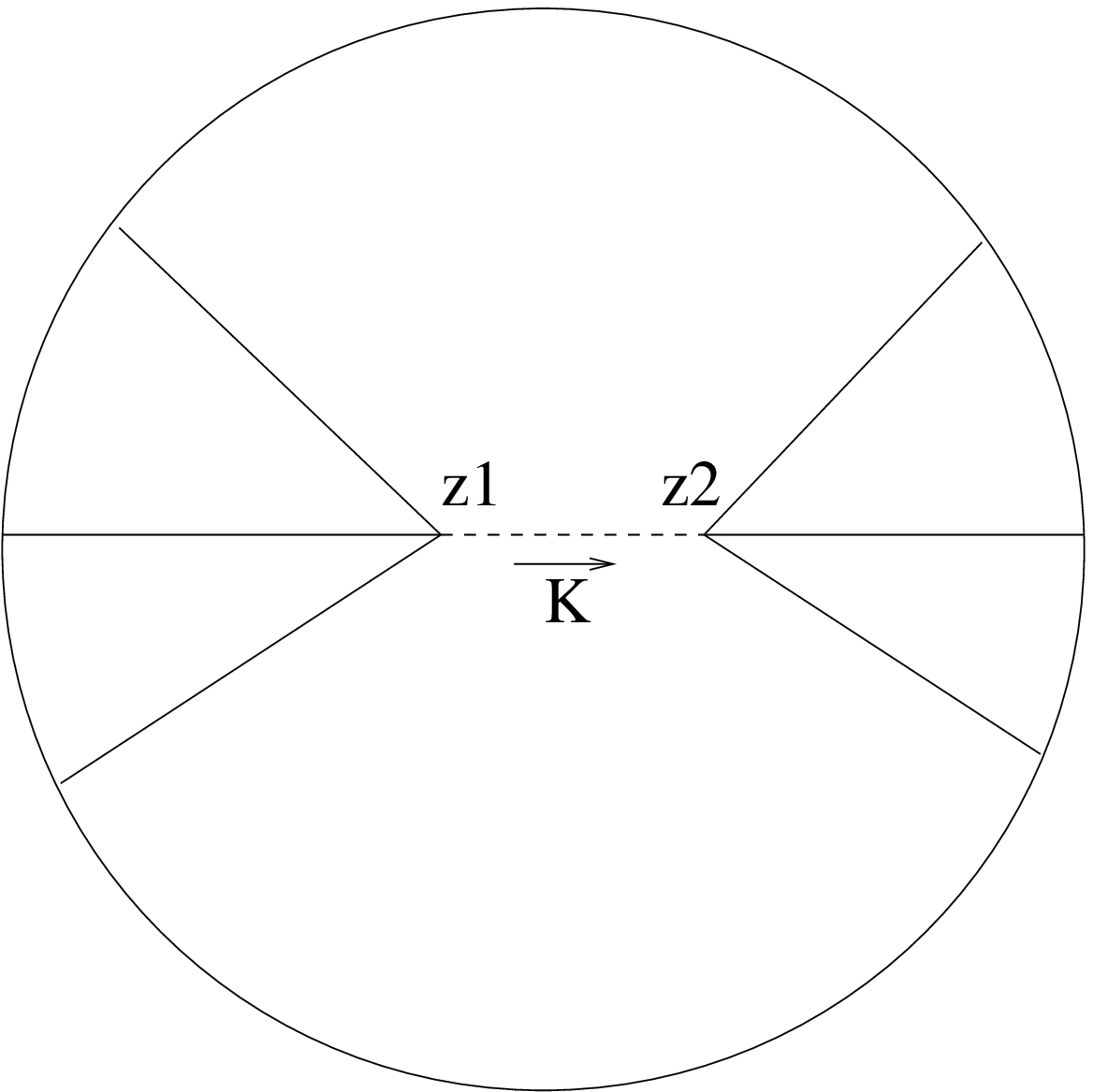}
\label{figadsexchange}
\caption{Exchange Interaction in AdS}
}
Consider a term with one propagator, as shown in the Fig. \ref{figadsexchange}, that runs between contact terms with $n_l$ lines on the left and $n_r$ lines on the right
This diagram can schematically be written 
\begin{equation}
\begin{split}
\label{contactwithprops}
&\int L^{i}_{j}( z_1, \vect{K}) G_{i l}^{j k}(z_1, z_2, \vect{K}) R^{l}_{k}(z_2, \vect{K}) {\sqrt{-g(z_1)}} \sqrt{-g(z_2)} d z_1 d z_2 \\ &= \int A_{l}^{k} (z_2, \vect{K}) R^{l}_{k} (z_2, \vect{K}) \sqrt{-g(z_2)} d z_2,
\end{split}
\end{equation}
where $L$ is the term from the left contact interaction, $R$ is the term from
the right contact interaction, $\vect{K}$ is the momentum running through
the propagator (we have Fourier transformed the propagator in the boundary
directions) and we have defined
\begin{equation}
\label{Adef}
 A_{l}^{k} (z_2, \vect{K}) \equiv
\int L^{i}_{j}(z_1, \vect{K}) G_{i l}^{j k}(z_1, z_2, \vect{K})  {\sqrt{-g(z_1)}} d z_1.
\end{equation}

Since $G$ is a Green's function we know that $A$ satisfies an ordinary
differential equation of the form:
\begin{equation}
\label{ode}
D^{i l}_{j k}(\vect{K}, z_2) A_{l}^{k}(z_2,\vect{K}) =  L^{i}_{j}( z_2, \vect{K}),
\end{equation}
where $D$ is a second order differential operator whose exact form is available easily\footnote{See for example, Equations (2.42), (2.37) and (2.32) in section 2.3:  ``Review of Perturbation Theory'' in \cite{Raju:2011ed}). 
$D$ can be read off from the quadratic part of the action.} but will not be important to us.
The argument that led to \eqref{contactvertexscaling}, but without adding
the contribution of the $\sqrt{-g}$ factor, now tells us that for large values of $z_2$ we have the scaling
\begin{equation}
 L^{i}_{j}(z_2, \vect{K})  \underset{z_2 \rightarrow \infty}{\longrightarrow} \Big(\prod_{q \in L} \norm{k^q} \Big)^{d - 1 \over 2} (z_2)^{n_l {d - 1 \over 2} + 2} {\cal L}^{i}_{j}(z_2, \vect{K}),
\end{equation}
where ${\cal L}^{i}_{j}$ is the corresponding interaction in flat space,
and the product over $q$ runs over all the momenta that appear on the 
left. Also for large $z_2$, we can verify that the differential operator scales like  $D(\vect{K},z_2) \sim z_2^2 {\cal D}(\vect{K},z_2)$, where ${\cal D}$ is the corresponding differential operator in flat space.

This means that for large $z_2$, $A$ must scale like
\begin{equation}
 A^{i}_{j}(z_2, \vect{K})  \underset{z_2 \rightarrow \infty}{\longrightarrow} \Big(\prod_{q \in L} \norm{k^q} \Big)^{d - 1 \over 2} (z_2)^{n_l {d - 1 \over 2}} {\cal A}^{i}_{j}(z_2, \vect{K}),
\end{equation}
where ${\cal A}$ is the quantity corresponding to \eqref{Adef} in flat space. Consequently (after using the scaling of $R$)  the final integrand inside \eqref{contactwithprops} must scale
like:
\begin{equation}
A_{l}^{k} (z_2, \vect{K}) R^{l}_{k} (z_2) \sqrt{-g(z_2)}  \underset{z_2 \rightarrow \infty}{\longrightarrow}  \Big(\prod_{q=1}^n \norm{k^q} \Big)^{d - 1 \over 2} z_2^{(n_l + n_r){d - 1 \over 2} - (d - 1)} {\cal A}_{l}^{k} (z_2, \vect{K}) {\cal R}^{l}_{k} (z_2),
\end{equation}
where the product over $q$ now runs over all momenta.

Since the location of the pole is governed by the behaviour of \eqref{contactwithprops} at large $z_2$, we are done and we get the pole we need including
the $\Gamma$ function from the scaling above.

\subsubsection{Exchange Interactions: Direct Integral Argument \label{secdirectintegral}}
We now give a second argument that is more direct and also sheds light
on  the 
exact analytic continuation that is required to observe this pole.  The relationship between contact interactions, which we derived above, evidently
holds diagram by diagram, with the flat-space diagrams evaluated in 
axial gauge. Consider the diagram \eqref{figadsexchange}, which is given by the expression \eqref{contactwithprops}. Consider
doing the integrals over both $z_1$ and $z_2$, but leaving the integral in $p$, which occurs in \eqref{gravitypropagator}, undone. Now, the ordinary
Bessel function that occurs in \eqref{gravitypropagator} has an asymptotic form that is given by:
\be
\label{besselasymptotic}
J_{d \over 2}(p z_1) \underset{z \rightarrow \infty}{\longrightarrow} \sqrt{2 \over \pi} \sin\left(p z_1 -({d + 1 \over 2}) {\pi \over 2} \right) {1 \over \sqrt{p z}}.
\ee
Repeating the argument for contact interactions above, and defining
\be
\label{etletrdef}
E_{T_L} = \sum_{q \in L} \norm{k^q}; \quad E_{T_R} = \sum_{q \in R} \norm{k^q},
\ee
we find that the integral over $z_1$ and $z_2$ gives
\begin{align}
\label{texchangeline1}
&T_{\text{ex}} =  {\Gamma(\alpha_{\text{gr}}^0(n_l + 1)) \Gamma(\alpha_{\text{gr}}^0(n_r + 1)) \left( \prod_{q=1}^n \norm{k^q} \right)^{d - 1 \over 2} \over 2 \pi} \\ 
\label{texchangeline2}  &\int d p\,  \Bigg[ \Big({e^{-i \pi (d + 1)  \over 4}  {\cal L}^{i}_{j}(\vect{K}, p)  \over (i p +  E_{T_L})^{\alpha_{\text{gr}}^0(n_l+1)}} -  {e^{i \pi (d + 1)  \over 4}  {\cal L}^{i}_{j}(\vect{K}, -p) \over (-i p +  E_{T_L} )^{\alpha_{\text{gr}}^0(n_l+1)}}\Big)    \\ \label{texchangeline3}
& 
 {\cal G}_{i l}^{j k}(p, \vect{K}) \Big({e^{-i \pi (d + 1)  \over 4}  {\cal R}^{l}_{k}(-\vect{K}, p)  \over (i p +  E_{T_R})^{\alpha_{\text{gr}}^0(n_r+1)}} -  {e^{i \pi (d + 1)   \over 4} {\cal R}^{l}_{k}(-\vect{K}, -p)  \over (-i p + E_{T_R} )^{\alpha_{\text{gr}}^0(n_r+1)}}\Big) + \ldots  \Bigg], 
\end{align}
where ${\cal G}$ is the flat-space graviton propagator in axial gauge, but Fourier transformed so that it is a function of the radial momentum $p$, 
rather than the radial coordinate. Similarly ${\cal L}$ and ${\cal R}$ have been Fourier transformed, and depend on the exchanged $d$-momentum $K$, and the  radial momentum $p$ rather then the radial coordinates $z_1$ and $z_2$ as in \eqref{contactwithprops}. The $\ldots$ indicate terms that have lower order poles in $(i p + E_{T_L})$ and $(-i p + E_{T_R})$. These will eventually give lower order poles in $E_T$

As we mentioned above, choosing the physical signs for the norms of the momenta while doing the $z$-integrals leads to a situation where  $(i E_{T_L})$ 
and $(i E_{T_R})$ are both in the first quadrant.  Note that the pre-factor of ${1 \over 2 \pi}$ comes about by multiplying the pre-factors in $\eqref{besselasymptotic}$ and accounting for the ${1 \over 2}$ in the $\sin$-function. Moreover, note that the product over $\norm{k^q}$ that appears in the pre-factor does not include a factor of $p$ because the Bessel function that appears in the propagator, and has the asymptotics \eqref{besselasymptotic}, is normalized differently from the bulk to boundary propagators described in \eqref{solgravfree}. 

\FIGURE{
\scalebox{0.4}{\input{contourpinch.pstex_t}}
\caption{Analytically continuing the poles along the dashed line pinches the contour}
\label{figcontourpinch}}

The contour of the $p$-integral runs from $0$ to $\infty$ and the integrand has at least four poles, that are shown in Fig. \ref{figcontourpinch}. Now, let us start analytically continuing the values of the norms of the momenta as shown by the dashed lines on left hand side of Fig. \ref{figcontourpinch}. When we reach a point where ${\text{Im}} \left( i E_{T_R} \right) = 0$, we can deform the $p$ contour upward to avoid the singularity. This defines an analytic continuation of $T_{\text{ex}}$. However, eventually we reach a point where the contour gets pinched between the poles as shown on the right hand side of Fig. \ref{figcontourpinch}. At this point $T_{\text{ex}}$ develops a singularity, since we cannot deform the contour any more. (See \cite{eden1966asm} for a nice discussion of singularities of complex integrals.)  

This singularity occurs when we take the first term inside the bracket in \eqref{texchangeline2}, which has a pole at $p = i E_{T_L}$ and multiply with the second term inside the bracket in  \eqref{texchangeline3}, which has a pole at $p = -i E_{T_R}$. This singularity is itself a pole, and we can determine the behaviour
near the singularity by evaluating the residue of the integrand in \eqref{texchangeline1} at $p = i E_{T_L}$ or at $p = -i E_{T_R}$. We find that
\be
T_{\text{ex}} \underset{E_T \rightarrow 0}{\longrightarrow}  {\Gamma(\alpha_{\text{gr}}^0(n)) \left( \prod_{q} \norm{k^q} \right)^{d - 1} \over \left(E_T\right)^{\alpha_{\text{gr}}^0(n)}}  {\cal L}^{i}_{j}(\vect{K}, E_{T_L})  {\cal G}_{i l}^{j k}(E_{T_L}, \vect{K}) {\cal R}^{l}_{k}(-\vect{K}, E_{T_R}).
\ee
The right hand side is just the value of the exchange diagram in flat space. So this leads exactly to the flat space limit indicated above.

\subsection{Flat Space Limit for Loop Amplitudes}
We would now like to generalize the flat space limit described above for
tree amplitudes to loop amplitudes. In this section, we will show that
\be
\label{flatspacestressloop}
M(\vect{e^1}, \vect{\tilde{k}^1}, \ldots \vect{e^n}, \vect{\tilde{k}^n}) = \lim_{E_T \rightarrow  0} {(E_T)^{\alpha_{\text{gr}}^l(n)}  \over \left(\prod_{m=1}^n \norm{k^m} \right)^{d - 1 \over 2} \Gamma(\alpha_{\text{gr}}^l) } T(\vect{e^1}, \vect{k^1}, \ldots \vect{e^n}, \vect{k^n}),
\ee
with 
\be
\label{alphagrl}
\alpha_{\text{gr}}^l(n) = {({n \over 2} - 1 + l)(d - 1) + 1},
\ee 
The equivalence under dilatations of the two sides above is a little more subtle. First we should note that both sides are UV-divergent within effective field theory. So we should properly understand the relation \eqref{flatspacestressloop} within dimensional regularization. Now, the flat space graviton $d+1$ dimensional scattering amplitude scales as $M \rightarrow \lambda^{2+l(d-1)} M$ under $\vect{k} \rightarrow \lambda \vect{k}$; this is precisely accounted for by the additional $l (d - 1)$ in $\alpha_{\text{gr}}^l$.

For current-correlators, we have a similar relation:
\be
M(\vect{\ep^1}, \vect{\tilde{k}^1}, \ldots \vect{\ep^n}, \vect{\tilde{k}^n}) = \lim_{E_T \rightarrow  0} {(E_T)^{\alpha_{\text{gl}}^l(n)}  \over \left(\prod \norm{k^m} \right)^{d - 3 \over 2} \Gamma(\alpha_{\text{gl}}^l) } T(\vect{\ep^1}, \vect{k^1}, \ldots \vect{\ep^n}, \vect{k^n}),
\ee
with
\be
\alpha_{\text{gl}}^l(n) = {({n \over 2} - 1 + l)(d - 3) + 1}.
\ee 
We will prove the relation between stress-tensor correlators and graviton amplitudes below, since the current-correlator $\leftrightarrow$ gluon-amplitude argument is almost identical.

\begin{figure} 
\begin{center}
\includegraphics[height=5cm]{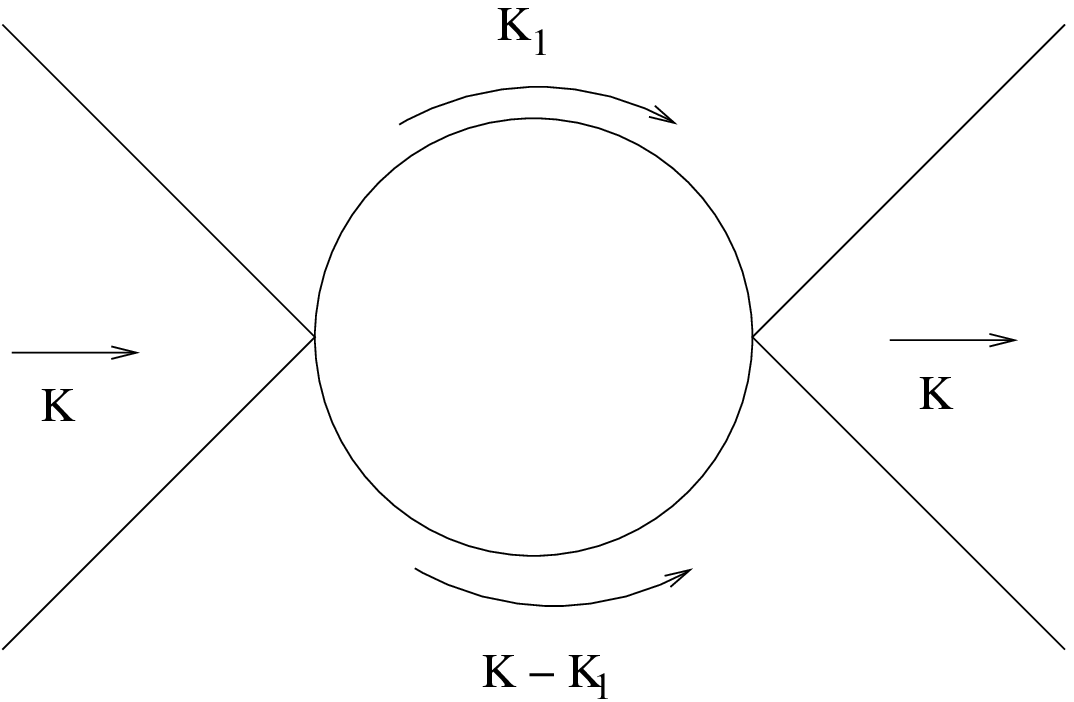}
\caption{One loop AdS diagram}
\label{figadsoneloop}
\end{center} 
\end{figure}

The $p$-integral argument above helps us make this generalization. Consider a loop diagram such as the one shown in figure \eqref{figadsoneloop}. 
This diagram can be written as
\begin{equation}
\begin{split}
\label{oneloopwithprops}
T_{1 \ell} = \int  &{\sqrt{-g(z_1)}} \sqrt{-g(z_2)} d z_1 d z_2 d^3 \vect{K}_1  \\ \Big[&L^{i_1 i_2}_{j_1 j_2}( z_1, \vect{K}) G_{i_1 l_1}^{j_1 k_1}(z_1, z_2, \vect{K}_1) G_{i_2 l_2}^{j_2 k_2}(z_1, z_2, \vect{K} - \vect{K}_1) R^{l_1 l_2}_{k_1 k_2}(z_2, -\vect{K})\Big].
\end{split}
\end{equation}
The key point is that we get the {\em product} of two (or more, if a higher-loop diagram is involved) bulk-bulk propagators. 
However, we can do 
the $z$-integrals to leave us with two integrals over radial momenta $p_1$ and $p_2$ and one
integral over the loop $d$-momentum:
\begin{align}
\label{toneline1} 
T_{1 \ell} = \int_{0}^{\infty} d p_1 \int_0^{\infty} &d p_2 \int {d^d \vect{K}_1 \over (2 \pi)^d} \Bigg[{\Gamma(\alpha_{\text{gr}}^0(n_l + 2)) \Gamma(\alpha_{\text{gr}}^0(n_r + 2)) \left( \prod_{q=1}^n \norm{k^q} \right)^{d - 1 \over 2} \over 4 \pi^2} \\ \label{toneline2}
 &\times  \Big\{ \sum_{s_i = \pm 1}  {e^{(s_1 + s_2) \pi {d + 1 \over 4}}  {\cal L}^{i_1 i_2}_{j_1 j_2}(\vect{K}, s_1 p_1, s_2 p_2) \over \big( i s_1 p_1 + i s_2 p_2 + E_{T_L}\big)^{\alpha_{\text{gr}}^0(n_l + 2)}} \Big\} {\cal G}_{i_1 l_1}^{j_1 k_1}(p_1, \vect{K}_1)  \\ \label{toneline3}
 &\times {\cal G}_{i_2 l_2}^{j_2 k_2}(p_2, \vect{K} - \vect{K}_1) \Big\{\sum_{r_i = \pm 1}    {e^{(r_1 + r_2) \pi {d + 1 \over 4}} {\cal R}^{l_1 l_2}_{k_1 k_2}(r_1 p_1, r_2 p_2, -\vect{K}) \over \big( i r_1 p_1 + i r_2 p_2 + E_{T_R}\big)^{\alpha_{\text{gr}}^0(n_r + 2)}}  \Big\} \Bigg].
\end{align}
The curly brackets on \eqref{toneline2} and \eqref{toneline3} come from
using the argument for the large-z scaling of contact interactions shown above. This expression is very similar to the expression in \eqref{texchangeline1} except that there are a total of 16 terms. We have introduced the 
compact variables $s_1, s_2, r_1, r_2$ that can each take the values $\pm 1$. 

Now let us consider doing the integral over $p_2$ first. As we mentioned
above this integral has at least 16 singularities that are all manifest
in the expression above. Now, recall that we start with $i E_{T_L}$ and $i E_{T_R}$ in the 
first quadrant. Since $p_1 \in (0, \infty)$ we see that for $p_1 < \text{Re}(i E_{T_L})$ the singularity corresponding to $p_2 = -p_1 + i E_{T_L}$ 
is also in the first quadrant. Now, we analytically continue $E_{T_R}$ exactly as shown in Fig. \ref{figcontourpinch}. More specifically, by flipping the signs of some of the $\norm{k^q}$ and then varying the values of the momenta, we get $-i E_{T_R}$ to the third quadrant, and then continue it upward
till it collides with $i E_{T_L}$. 

This leads a singularity at $E_T = 0$ in the integral when the singularities in the integrand at  $i p_2 + i p_1 + E_{T_L} = 0$, and $-i p_2 - i p_1 + E_{T_R} = 0$ collide and pinch the $p_2$ contour. On the other hand, for 
$p_1 > \text{Re}(i E_{T_L})$, we get a singularity in the integral at $E_T = 0$, when the singularities in the integrand  at $-i p_2 + i p_1 + E_{T_L} = 0$ and $i p_2 - i p_1 + E_{T_R} = 0$ collide. 

We also get a singularity in the integral at $E_T = 0$ when the singularities in the integrand at  $i p_2 - i p_1 + E_{T_L} = 0$ and $-i p_2 + i p_1 + E_{T_R} = 0$ collide. These combinations are all summarized in Table \ref{tabsingularities}.
\TABLE{
\label{tabsingularities}
\begin{tabular}{|c|c|}
\hline Condition & Colliding Singularities \\ \hline
$p_1 < \text{Re}(i E_{T_L})$ & $i p_2 + i p_1 + E_{T_L} = 0$ and $ -i p_2 - i p_1 + E_{T_R} = 0$ \\
$p_1 > \text{Re}( i E_{T_L})$ & $-i p_2 + i p_1 + E_{T_L} = 0$ and $i p_2 - i p_1 + E_{T_R} = 0$ \\
All $p_1$ & $i p_2 - i p_1 + E_{T_L} = 0$ and $-i p_2 + i p_1 + E_{T_R} = 0$. \\ \hline
\end{tabular}
\caption{Colliding singularities in the integrand give rise to a $E_T = 0$ singularity in the integral.}
}

Taking one of the contributions from the first two lines of Table \ref{tabsingularities} and the contribution from the third line gives us 
the following answer
\begin{align}
\label{toneline4} 
T_{1 \ell} = \int_0^{\infty} & {d p_1 \over 2 \pi} \int {d^d \vect{K}_1 \over (2 \pi)^d} \Bigg[{\Gamma(\alpha_{\text{gr}}^1(n)) \left( \prod_{q=1}^n \norm{k^q} \right)^{d - 1 \over 2} } \left({1 \over E_T} \right)^{\alpha_{\text{gr}}^1(n_l + n_r)}\\ \label{toneline5}
 & \Big\{ {\cal L}^{i_1 i_2}_{j_1 j_2}(\vect{K}, -p_1, p_2 = i E_{T_L} - p_1) {\cal G}_{i_1 l_1}^{j_1 k_1}(p_1,  \vect{K}_1)  \\ \label{toneline6}
 & {\cal G}_{i_2 l_2}^{j_2 k_2}( p_2= i E_{T_L} - p_1, \vect{K} - \vect{K}_1) {\cal R}^{l_1 l_2}_{k_1 k_2}(p_1, -p_2 = i E_{T_R} + p_1, \vect{K}) \Big\} \\ \label{toneline7} &+ p_1 \rightarrow -p_1  + \ldots \Bigg].
\end{align}
Here $\ldots$ are terms that have lower order singularities in $E_{T}$. However, the $p_1 \rightarrow -p_1$ interchange in \eqref{toneline7} is exactly what we need to convert the integral over $p_1$ from $(0,\infty)$ to $(-\infty, \infty)$. We can now combine the integral over $\vect{K}_1$ and the integral over $p_1$ into a single $d+1$-dimensional loop-integral, which is what occurs in the flat-space amplitude. This leads to the result \eqref{flatspacestressloop} with $l = 1$. 

We can show the generalization to arbitrary $l$ through induction. Consider a $l$-loop diagram,
which is made up of a $m_l$-loop diagram on the left, a $m_r$-loop diagram
on the right and let us focus on the $l - m_l - m_r$ loops in the middle. We can write this diagram in the form \eqref{toneline1}, although the exponents of the singularities associated with ${\cal L}$ and ${\cal R}$ will now be $\alpha^{m_l}_{\text{gr}}(n_l + l - m_l - m_r + 1)$ and $\alpha^{m_r}_{\text{gr}}(n_r + l - m_l - m_r + 1)$. To obtain the pole in $E_T$, we can make an argument similar to the one above. The key identity that we need is that:
\be
\alpha_{\text{gr}}^l(n_l + n_r) = \alpha_{\text{gr}}^{m_l}(n_l + l - m_l - m_r + 1 ) + \alpha_{\text{gr}}^{m_r}(n_r + l - m_l - m_r + 1) - 1,
\ee
which holds irrespective of the values of $m_l$ and $m_r$. 

\subsection{Flat Space Limit of the Recursion Relations \label{secflatspaceofrecurs}}
We wish to prove that our recursion relations have the right flat space limit. We will do this by induction. The recursion relations take three-point
amplitudes as an input, and then generate higher point amplitudes. The
three-point amplitudes need to be computed directly through a bulk AdS
computation, or some other method, and by the argument of section \ref{secflatspace}, they automatically have the correct flat space limit. In fact
this can be seen more concretely in the results for three-point functions obtained in \cite{Maldacena:2011nz}. 

Now, to make the inductive argument,  let us assume that all $m$-point amplitudes with $m$ smaller than some given $n$ have the right flat space limit. If we now compute a higher point amplitude using \eqref{stressrecurs},
our assumption states that both the $q$-point amplitude and the $n-q$ point amplitude in ${\cal T}^2$ have the right flat space limit. In particular, this means that \eqref{stressrecurs} involves a term:
\begin{align}
\label{stressrecursflat1}
T(\vect{e^1}, \vect{k^1}, \ldots \vect{e^n}, \vect{k^n}) &= {\cal B} + \sum_{\{\pi\},\vect{e^{m'}}. \pm}\int  {i {\cal T}_f^2 \over p^2 + (\sum_{o=1}^{m_l} \vect{k^{\pi_o}})^2}  {d p^2 \over 2} {w^{\mp}(p) \over w^{\pm}(p) - w^{\mp}(p)} + \ldots, \\
\label{stressrecursflat2}
{\cal T}_f^2 \equiv  \prod_{o=1}^n \norm{k^o} &\left( \Gamma(q+1)   \sum_{s = \pm 1} {M(\vect{e^{\pi_1}} , \vect{\tilde{k}^{\pi_1}}(p),  \ldots \vect{e^{q'}}, \vect{\tilde{k}^{q'_s}}) \over (E_{T_L} +  i s p)^{\alpha_{gr}^0(q+1)}} \right) \\ \times  &\left(\Gamma(n-q+1) \sum_{r = \pm 1}{  M(\vect{e^{q'}}, \{-\vect{\tilde{k}^{q'_r}} \ldots \vect{e^n}, \vect{\tilde{k}^{\pi_n}}(p)) \over (E_{T_R} +  i r p)^{\alpha_{gr}^0(n-q+1)}}  \right). 
\end{align}
Here, $M$ is the flat space amplitude as in section \ref{secflatrecurs}, $E_{T_L}$ and $E_{T_R}$ have the same definition as \eqref{etletrdef} and  the $\ldots$ in \eqref{stressrecursflat1} indicate terms that will give a lower order pole in $E_T$ after the $p$-integral is done. The symbols $\vect{\tilde{k}^m}$ have the same meaning as in \eqref{tildekdef} and 
\be
\vect{\tilde{k}^{q'_s}} \equiv \{\vect{k}^{q'}, i s p\} ; \quad \vect{\tilde{k}^{q'_r}} \equiv \{\vect{k}^{q'}, i r p\}.
\ee

Now using exactly the same argument as section \ref{secdirectintegral}, we
see that the $n$-point correlator has a term:
\begin{equation}
\label{stressrecurstogravrecurs}
\begin{split}
&T(\vect{e^1}, \vect{k^1}, \ldots \vect{e^n}, \vect{k^n}) = {\Gamma(n)  \prod_{o=1}^n \norm{k^o} \over E_T^{\alpha_{gr}^0(n)}} \sum_{\{\pi\},\vect{e^{m'}}. \pm}  {i {\cal M}^2 \over p^2 + (\sum_{o=1}^{m_l} k_{\pi_o})^2}   {w^{\mp}(p) \over w^{\pm}(p) - w^{\mp}(p)} + \ldots, \\
&{\cal M}^2 \equiv  {M(\vect{e^{\pi_1}}, \vect{k^{\pi_1}}(p),  \ldots \vect{e^{q'}}, \vect{k^{q'}}) M(\vect{e^{q'}}, -\vect{k^{q'}},\ldots \vect{e^n}, \vect{k^{\pi_n}}(p)) }.
\end{split}
\end{equation}
where $\ldots$ are terms that have lower order singularities in $E_T$.

However, we see the coefficient of the highest order singularity at $E_T = 0$ is just what appears in the flat-space recursion relations \eqref{gravityrecurs}, which generate the flat space scattering amplitudes. This proves that the recursion relations \eqref{stressrecurs} have the correct flat space limit.

\section{Conclusions}
There are two main results in this paper.  Our first result has to do with a new set recursion relations for correlation functions of the stress tensor and conserved currents in AdS/CFT. To find these recursion relations, we first developed a new set of recursion relations for  graviton and gluon tree amplitudes in flat space. These are presented in Equations \eqref{ymrecurs} and \eqref{gravityrecurs}.  We then generalized these recursion relations to anti-de Sitter space: these generalizations
are presented in Equations \eqref{currecurs} and \eqref{stressrecurs}.   Our new recursion relations rely on extending each momentum by its polarization-vector. These relations have an advantage over the BCFW-like relations 
derived in  \cite{Raju:2011ed,Raju:2010by} since they are valid for AdS$_4$/CFT$_3$. In higher dimensions --- while they give rise to more terms than the BCFW relations --- they involve less stringent conditions on the polarizations than the conditions enumerated in \cite{Raju:2011ed,Raju:2010by}. Moreover, they can be used to explicitly maintain crossing symmetry. 

Our second main result in this paper was a new method of extracting flat space S-matrix elements from AdS/CFT correlators. In particular we showed that given a stress tensor correlator in a conformal field theory with a bulk pure gravity dual, one could recover the $(d+1)$ dimensional graviton amplitude in flat space using \eqref{flatspacestressloop}. This flat space limit is valid beyond tree level, at any fixed order in perturbation theory. 

We then showed that our recursion relations automatically generated answers that had the correct flat space limit. This is a powerful consistency check on their validity. 

In an accompanying paper, we have shown how these results may be used in a concrete setting. In \cite{Raju:2012zs},  we used the recursion relations to obtain explicit answers for four point correlation functions of the stress tensor in AdS$_4$/CFT$_3$ and then checked these answers by verifying that, in the flat space limit, they reduce to the famous formulas for four point graviton amplitudes.

We should mention that although our explicit results for the flat space limit were derived for the case of pure gravity and Yang-Mills theory it is clear how  we must proceed in the presence of other kinds of interactions. For example,
in the presence of an $R^3$ interaction we would just get a factor
of $z^6$ instead of a $z^2$ in \eqref{conformaltransricci}. This would give rise to higher order poles, which is indeed what is observed in the computations with a Weyl-cubed action in \cite{Maldacena:2011nz}. 

If we have both an $R$ and a $R^3$ term in the action, we can still use our flat space limit provided these terms are multiplied by adjustable parameters. For example, in string theory on AdS$_5 \times S^5$, there are higher derivative
terms in the effective action that are suppressed by factors of ${1 \over \lambda}$. So if we could somehow compute stress tensor correlators in the strongly coupled ${\cal N} = 4$ SYM theory then to compare the results with
the prescription given in this paper, we would need to expand the answer {\em both} in ${1 \over N}$ and ${1 \over \lambda}$. The leading term in this expansion (both in ${1 \over N}$ and ${1 \over \lambda}$) is reproduced by tree-level gravity in AdS$_5 \times S^5$ and should have the flat space limit indicated above. Furthermore, if we stick to leading order in ${1 \over N}$ then the higher 
order terms in ${1 \over \lambda}$ will have higher order poles whose residues will reproduce the corrections to graviton amplitudes by higher derivative corrections in flat space string theory. 

On the other hand, it is unclear how this method should be applied to theories like the
Vasiliev theory \cite{Vasiliev:2003ev,Vasiliev:1999ba,Bekaert:2005vh,Iazeolla:2008bp,Campoleoni:2009je}, where higher derivative terms are not suppressed by any
parameter. In this case, we might get arbitrarily high order poles at any given order in the ${1 \over N}$ expansion. It would be nice to see if our flat space limit can be generalized to apply to that case also.

Vasiliev-type theories also seem to present obstacles to the recursion 
relations because it seems hard to control the behaviour of the correlator
at $w = \infty$. On the other hand, given that the BCFW recursion
relations can be generalized to string theory \cite{Boels:2008fc,Cheung:2010vn,Boels:2010bv}, we could hope that some generalization of these new recursion relations might work for higher spin theories as well.  This would be very useful since computations
with higher spins  are even harder than gravity computations. More ambitiously, since these recursion relations determine all correlators starting with just the three point transition amplitude, it would be nice to explore whether it is possible to use these techniques
to demonstrate the equivalence of the Vasiliev theory and the O(N) vector
model to all orders \cite{Klebanov:2002ja,Giombi:2010vg,Giombi:2011ya}.

\section*{Acknowledgments}
I am grateful to Juan Maldacena and Guilherme Pimentel for collaboration in the early stages of this work. I would like to thank Sayantani Bhattacharyya, Bobby Ezhuthachan, Rajesh Gopakumar, Shiraz Minwalla, Kyriakos Papadodimas, Sandip Trivedi and Ashoke Sen for useful discussions. This work was primarily supported by a Ramanujan Fellowship of the Department of Science and Technology. I would also like to acknowledge the support of the Harvard University Department of Physics. I would like to thank the Institute for Advanced Study (Princeton), the Institute of Mathematical Sciences (Chennai), the Chennai Mathematical Institute, Delhi University and the Tata Institute of Fundamental Research (Mumbai) for their hospitality while this work was being completed.
\appendix
\section*{Appendix}
\section{Difficulties with BCFW in AdS$_4$/CFT$_3$}
In this appendix we briefly describe the difficulties involved in generalizing
the BCFW recursion relations to the computation of correlation
functions in AdS$_4$/CFT$_3$. It is entirely possible that these difficulties are surmountable and we present 
this analysis here in the hope that a reader of this paper will find a way to improve it. In fact the development of BCFW relations for AdS$_4$/CFT$_3$ would be quite valuable. The BCFW-recursion relations involve fewer terms than \eqref{stressrecurs} because the sum over partitions is limited to partitions in which one chosen momentum appears on the left, and another chosen momentum appears on the right. This means that such relations are likely to more directly lead to compact expressions for final answers.

The standard BCFW relations rely on finding a null vector $\vect{q}$ that is orthogonal to two given momenta $\vect{k^1}$ and $\vect{k^n}$ i.e. we must have
\be
\vect{q} \cdot \vect{q} = \vect{q} \cdot \vect{k^1} = \vect{q} \cdot \vect{k^n} = 0.
\ee

Given two arbitrary momenta --- $\vect{k^1}$ and $\vect{k^n}$ --- in three dimensions, there is no solution to this equation even if we allow $\vect{q}$ to become complex.  One solution to this problem, which works for scattering amplitudes that depend on massless momenta in 3 dimensions was developed in \cite{Gang:2010gy}. Here, we will try and generalize it to the computation 
of correlators, which can depend on arbitrary momenta. 

The idea is that given two vectors $\vect{k^1}$ and $\vect{k^n}$, we want
to ``rotate'' them in the plane, while keeping their sum constant. We will allow the 
``angle of rotation'' to take complex values. 

Let us  define 
\begin{equation}
 k^m_{\alpha \dot{\alpha}} = |\vect{k^m}| \sigma^0_{\alpha \dot{\alpha}} + 
 k^m_i \sigma^i_{\alpha \dot{\alpha}},
\end{equation}
for $m=1$ or $m=n$. We also write
\begin{equation}
k^1_{\alpha \dot{\alpha}} = \la^1_{\alpha} \lb^1_{\dot{\alpha}}, \quad k^n_{\alpha \dot{\alpha}} = \la^n_{\alpha} \lb^n_{\dot{\alpha}},
\end{equation}
where $\la^1, \la^n, \lb^1, \lb^n$ are two component spinors.

Then the following rotation has the properties that we want:
\begin{equation}
R = \exp\{-i {\theta \vect{\sigma} \cdot (\vect{k^1} + \vect{k^n}) \over |\vect{k^1} + \vect{k^n}|}\}; \quad R^{-1} = \exp\{+i {\theta \vect{\sigma} \cdot (\vect{k^1} + \vect{k^n}) \over |\vect{k^1} + \vect{k^n}|}\}.
\end{equation}
Under which the spinors transform as
\begin{equation}
\label{spinortrans}
\la_m \longrightarrow R \la_m, \quad \lb_m \longrightarrow \lb_m R^{-1}.
\end{equation}
In particular, with $\hat{\vect{n}} \equiv  {(\vect{k^1} + \vect{k^n}) \over |\vect{k^1} + \vect{k^n}|}$, we have 
\begin{equation}
R = \cos {\theta \over 2} - i \vect{\sigma} \cdot \hat{\vect{n}} \sin{\theta \over 2} 
= {1 \over 2} \left[(x + 1/x) - \vect{\sigma} \cdot \hat{\vect{n}} (x - 1/x)\right],
\end{equation}
with $x \equiv e^{i \theta \over 2}$. However, we do not need to restrict
to $|x| = 1$ and can consider this rotation to be an arbitrary function of $x$.

Now, since the norm of both vectors $\vect{k^1}$ and $\vect{k^n}$ is independent of $x$, it is clear that the correlator
can be written as an integral over a rational function of $x$. This integrand has poles when an intermediate propagator goes on shell. However, it also has potential poles at $x = 0$ and at $x = \infty$.

Let us choose a coordinate system to gain some intuition for what happens under this extension. In particular, we choose
\begin{equation}
\vect{k^1} = (0, 1, \alpha), \quad \vect{k^n} = (0,-1,\beta).
\end{equation}
where we have rescaled coordinates so that the $y$-component of the vectors is
$1$, without loss of generality. 
(These expressions are written as three dimensional expressions.)
Initially, these vectors are associated with spinors
\begin{equation}
\begin{split}
&{\la_1} = \left\{\sqrt{\alpha +\sqrt{\alpha ^2+1}},\frac{i}{\sqrt{\alpha
   +\sqrt{\alpha ^2+1}}}\right\},
 \quad {\lb_1} = \left\{\sqrt{\alpha +\sqrt{\alpha ^2+1}},-\frac{i}{\sqrt{\alpha
   +\sqrt{\alpha ^2+1}}}\right\}, \\
&{\la_n} = \left\{\sqrt{\beta +\sqrt{\beta ^2+1}},-\frac{i}{\sqrt{\beta
   +\sqrt{\beta ^2+1}}}\right\},\quad {\lb_n} = \left\{\sqrt{\beta +\sqrt{\beta ^2+1}},\frac{i}{\sqrt{\beta
   +\sqrt{\beta ^2+1}}}\right\}.
\end{split}
\end{equation}
As we make our rotation above, the momenta get transformed to
\begin{equation}
\begin{split}
&\vect{k^1}(x) = \left\{\frac{1}{2} i
   \left(x^2-\frac{1}{x^2}\right),\frac{-1}{2}
   \left(-x^2-\frac{1}{x^2}\right),\alpha \right\}, \\
&\vect{k^n}(x) = \left\{-\frac{1}{2} i
   \left(x^2-\frac{1}{x^2}\right),\frac{1}{2}
   \left(-x^2-\frac{1}{x^2}\right),\beta \right\},
\end{split}
\end{equation}
with associated negative helicity polarizations (obtained using the spinor
transformation rule \eqref{spinortrans}) that are
\begin{equation}
\begin{split}
\ep^-_1(x) &= \left\{-\frac{x^4+2 \alpha ^2+2 \alpha  \sqrt{\alpha ^2+1}+1}{2 x^2
   \left(\alpha +\sqrt{\alpha ^2+1}\right)},\frac{i \left(x^4-2 \alpha
   ^2-2 \alpha  \sqrt{\alpha ^2+1}-1\right)}{2 x^2 \left(\alpha
   +\sqrt{\alpha ^2+1}\right)},i\right\} \\ &= i  \left\{\frac{1}{2} i
   \left(\gamma_1 x^2+\frac{1}{\gamma_1 x^2}\right),\frac{1}{2}
   \left(\gamma_1 x^2-\frac{1}{\gamma_1 x^2}\right),1\right\},\\
\ep^-_n(x) &= \left\{-\frac{x^4+2 \beta ^2+2 \beta  \sqrt{\beta ^2+1}+1}{2 x^2
   \left(\beta +\sqrt{\beta ^2+1}\right)},\frac{i \left(x^4-2 \beta
   ^2-2 \beta  \sqrt{\beta ^2+1}-1\right)}{2 x^2 \left(\beta
   +\sqrt{\beta ^2+1}\right)},-i\right\} \\ &= -i  \left\{\frac{1}{2} i
   \left(\gamma_n x^2+\frac{1}{\gamma_n x^2}\right),\frac{1}{2}
   \left(\gamma_n x^2-\frac{1}{\gamma_n x^2}\right),1\right\}.
\end{split}
\end{equation}
The positive helicity polarizations are similar
\begin{equation}
\begin{split}
&\ep^+_1(x)= \left\{0,-\frac{\left(2 \alpha ^2+2 \sqrt{\alpha ^2+1} \alpha
   +1\right) x^4+1}{2 x^2 \left(\alpha +\sqrt{\alpha
   ^2+1}\right)},\frac{i \left(x^4 \left(2 \alpha ^2+2 \sqrt{\alpha
   ^2+1} \alpha +1\right)-1\right)}{2 x^2 \left(\alpha +\sqrt{\alpha^2+1}\right)},-i\right\} \\
&= i  \left\{\frac{1}{2} i
   \left({x^2 \over \gamma_1} +\frac{\gamma_1}{x^2}\right),\frac{1}{2}
   \left({x^2 \over \gamma_1}-\frac{\gamma_1}{x^2}\right),-1\right\},\\
&\ep^+_n(x)= \left\{0,-\frac{\left(2 \beta ^2+2 \sqrt{\beta ^2+1} \beta
   +1\right) x^4+1}{2 x^2 \left(\beta +\sqrt{\beta
   ^2+1}\right)},\frac{i \left(x^4 \left(2 \beta ^2+2 \sqrt{\beta
   ^2+1} \beta +1\right)-1\right)}{2 x^2 \left(\beta +\sqrt{\beta^2+1}\right)},i\right\} \\
&= -i  \left\{\frac{1}{2} i
   \left({x^2 \over \gamma_n} +\frac{\gamma_n}{x^2}\right),\frac{1}{2}
   \left({x^2 \over \gamma_n} -\frac{\gamma_n}{x^2}\right),-1\right\}.
\end{split}
\end{equation}

However, these polarization vectors blow up {\em both} at $x = 0$ and $x = \infty$. 
If we consider a gluon amplitude then naively we would expect that for large $x$, following the analysis of \cite{ArkaniHamed:2008yf}, that the amplitude would behave like
behave like:
\begin{equation}
T_4 \sim \ep^1_{i} \eta^{i j} x^2 \ep^n_{j}  + \ldots
\end{equation}
So, we might expect that $T_4 \sim {\rm O}(x^6)$, since both polarizations
grow like $x^2$. However, since $\vect{\ep^1} = \gamma_1 \vect{k^1} + {\rm O}(1)$ and similarly for $\vect{\ep^n}$, we can use the Ward identity twice to get rid of a factor of $x^4$. (More precisely the highest order terms in $x$ are fixed by
the contact terms that appear in the Ward identity.) However, this still leaves us with the scaling $T_4 \sim {\rm O}(x^2)$.  The same problem occurs at $x = 0$. 

We do not know how to compute these boundary terms in a simple way. Moreover, if we try and get rid of this problem by scaling the polarization vectors as we go to $x \rightarrow \infty$ and also as we go to $x \rightarrow 0$, we inevitably introduce a pole somewhere else in the complex plane with residues that
do not have any nice physical interpretation. For this reason, the naive approach to the BCFW recursion relations for AdS$_4$/CFT$_3$ runs into trouble. 

\bibliographystyle{JHEP}
\bibliography{references}

\end{document}

%% file: contourpinch.pstex_t
\begin{picture}(0,0)%
\includegraphics{contourpinch.pstex}%
\end{picture}%
\setlength{\unitlength}{3947sp}%
\begingroup\makeatletter\ifx\SetFigFont\undefined%
\gdef\SetFigFont#1#2#3#4#5{%
  \reset@font\fontsize{#1}{#2pt}%
  \fontfamily{#3}\fontseries{#4}\fontshape{#5}%
  \selectfont}%
\fi\endgroup%
\begin{picture}(14415,3669)(4189,-3418)
\put(7396,-1051){\makebox(0,0)[lb]{\smash{{\SetFigFont{14}{16.8}{\rmdefault}{\mddefault}{\updefault}{\color[rgb]{0,0,0}$i E_{TL}$}%
}}}}
\put(13456,-2506){\makebox(0,0)[lb]{\smash{{\SetFigFont{14}{16.8}{\rmdefault}{\mddefault}{\updefault}{\color[rgb]{0,0,0}$-i E_{TL}$}%
}}}}
\put(15841,-736){\makebox(0,0)[lb]{\smash{{\SetFigFont{14}{16.8}{\rmdefault}{\mddefault}{\updefault}{\color[rgb]{0,0,0}$i E_{TL}$}%
}}}}
\put(15901,-1501){\makebox(0,0)[lb]{\smash{{\SetFigFont{14}{16.8}{\rmdefault}{\mddefault}{\updefault}{\color[rgb]{0,0,0}$-i E_{TR}$}%
}}}}
\put(7366,-436){\makebox(0,0)[lb]{\smash{{\SetFigFont{14}{16.8}{\rmdefault}{\mddefault}{\updefault}{\color[rgb]{0,0,0}$i E_{TR}$}%
}}}}
\put(4471,-1921){\makebox(0,0)[lb]{\smash{{\SetFigFont{14}{16.8}{\rmdefault}{\mddefault}{\updefault}{\color[rgb]{0,0,0}$-i E_{TL}$}%
}}}}
\put(4366,-3076){\makebox(0,0)[lb]{\smash{{\SetFigFont{14}{16.8}{\rmdefault}{\mddefault}{\updefault}{\color[rgb]{0,0,0}$-i E_{TR}$}%
}}}}
\put(13441,-1831){\makebox(0,0)[lb]{\smash{{\SetFigFont{14}{16.8}{\rmdefault}{\mddefault}{\updefault}{\color[rgb]{0,0,0}$i E_{TR}$}%
}}}}
\end{picture}%

%% file: newrecursion.bbl
\providecommand{\href}[2]{#2}\begingroup\raggedright\begin{thebibliography}{10}

\bibitem{Maldacena:1997re}
J.~M. Maldacena, {\it {The Large N limit of superconformal field theories and
  supergravity}},  {\em Adv.Theor.Math.Phys.} {\bf 2} (1998) 231--252,
  [\href{http://xxx.lanl.gov/abs/hep-th/9711200}{{\tt hep-th/9711200}}].

\bibitem{Gubser:1998bc}
S.~Gubser, I.~R. Klebanov, and A.~M. Polyakov, {\it {Gauge theory correlators
  from noncritical string theory}},  {\em Phys.Lett.} {\bf B428} (1998)
  105--114, [\href{http://xxx.lanl.gov/abs/hep-th/9802109}{{\tt
  hep-th/9802109}}].

\bibitem{Witten:1998qj}
E.~Witten, {\it {Anti-de Sitter space and holography}},  {\em Adv. Theor. Math.
  Phys.} {\bf 2} (1998) 253--291,
  [\href{http://xxx.lanl.gov/abs/hep-th/9802150}{{\tt hep-th/9802150}}].

\bibitem{DeWitt:1967uc}
B.~S. DeWitt, {\it {Quantum theory of gravity. III. Applications of the
  covariant theory}},  {\em Phys. Rev.} {\bf 162} (1967) 1239--1256.

\bibitem{Parke:1986gb}
S.~J. Parke and T.~R. Taylor, {\it {An Amplitude for $n$ Gluon Scattering}},
  {\em Phys. Rev. Lett.} {\bf 56} (1986) 2459.

\bibitem{Berends:1988zp}
F.~A. Berends, W.~Giele, and H.~Kuijf, {\it {On relations between multi - gluon
  and multigraviton scattering}},  {\em Phys.Lett.} {\bf B211} (1988) 91.

\bibitem{Britto:2004ap}
R.~Britto, F.~Cachazo, and B.~Feng, {\it {New recursion relations for tree
  amplitudes of gluons}},  {\em Nucl. Phys.} {\bf B715} (2005) 499--522,
  [\href{http://xxx.lanl.gov/abs/hep-th/0412308}{{\tt hep-th/0412308}}].

\bibitem{Britto:2005fq}
R.~Britto, F.~Cachazo, B.~Feng, and E.~Witten, {\it {Direct proof of tree-level
  recursion relation in Yang- Mills theory}},  {\em Phys. Rev. Lett.} {\bf 94}
  (2005) 181602, [\href{http://xxx.lanl.gov/abs/hep-th/0501052}{{\tt
  hep-th/0501052}}].

\bibitem{Witten:2003nn}
E.~Witten, {\it {Perturbative gauge theory as a string theory in twistor
  space}},  {\em Commun. Math. Phys.} {\bf 252} (2004) 189--258,
  [\href{http://xxx.lanl.gov/abs/hep-th/0312171}{{\tt hep-th/0312171}}].

\bibitem{ArkaniHamed:2008gz}
N.~Arkani-Hamed, F.~Cachazo, and J.~Kaplan, {\it {What is the Simplest Quantum
  Field Theory?}},  {\em JHEP} {\bf 09} (2010) 016,
  [\href{http://xxx.lanl.gov/abs/0808.1446}{{\tt arXiv:0808.1446}}].

\bibitem{ArkaniHamed:2009dn}
N.~Arkani-Hamed, F.~Cachazo, C.~Cheung, and J.~Kaplan, {\it {A Duality For The
  S Matrix}},  {\em JHEP} {\bf 1003} (2010) 020,
  [\href{http://xxx.lanl.gov/abs/0907.5418}{{\tt arXiv:0907.5418}}].

\bibitem{Forde:2007mi}
D.~Forde, {\it {Direct extraction of one-loop integral coefficients}},  {\em
  Phys. Rev.} {\bf D75} (2007) 125019,
  [\href{http://xxx.lanl.gov/abs/0704.1835}{{\tt arXiv:0704.1835}}].

\bibitem{Mason:2008jy}
L.~Mason and D.~Skinner, {\it {Gravity, Twistors and the MHV Formalism}},  {\em
  Commun.Math.Phys.} {\bf 294} (2010) 827--862,
  [\href{http://xxx.lanl.gov/abs/0808.3907}{{\tt arXiv:0808.3907}}].

\bibitem{Spradlin:2008bu}
M.~Spradlin, A.~Volovich, and C.~Wen, {\it {Three Applications of a Bonus
  Relation for Gravity Amplitudes}},  {\em Phys.Lett.} {\bf B674} (2009)
  69--72, [\href{http://xxx.lanl.gov/abs/0812.4767}{{\tt arXiv:0812.4767}}].

\bibitem{Raju:2009yx}
S.~Raju, {\it {The Noncommutative S-Matrix}},  {\em JHEP} {\bf 06} (2009) 005,
  [\href{http://xxx.lanl.gov/abs/0903.0380}{{\tt arXiv:0903.0380}}].

\bibitem{Lal:2009gn}
S.~Lal and S.~Raju, {\it {The Next-to-Simplest Quantum Field Theories}},  {\em
  Phys. Rev.} {\bf D81} (2010) 105002,
  [\href{http://xxx.lanl.gov/abs/0910.0930}{{\tt arXiv:0910.0930}}].

\bibitem{Nguyen:2009jk}
D.~Nguyen, M.~Spradlin, A.~Volovich, and C.~Wen, {\it {The Tree Formula for MHV
  Graviton Amplitudes}},  \href{http://xxx.lanl.gov/abs/0907.2276}{{\tt
  arXiv:0907.2276}}.

\bibitem{Lal:2010qq}
S.~Lal and S.~Raju, {\it {Rational Terms in Theories with Matter}},  {\em JHEP}
  {\bf 08} (2010) 022, [\href{http://xxx.lanl.gov/abs/1003.5264}{{\tt
  arXiv:1003.5264}}].

\bibitem{Boels:2010nw}
R.~H. Boels, {\it {On BCFW shifts of integrands and integrals}},  {\em JHEP}
  {\bf 1011} (2010) 113, [\href{http://xxx.lanl.gov/abs/1008.3101}{{\tt
  arXiv:1008.3101}}].

\bibitem{Raju:2010by}
S.~Raju, {\it {BCFW for Witten Diagrams}},  {\em Phys. Rev. Lett.} {\bf 106}
  (2011) 091601, [\href{http://xxx.lanl.gov/abs/1011.0780}{{\tt
  arXiv:1011.0780}}].

\bibitem{Raju:2011ed}
S.~Raju, {\it {Recursion Relations for AdS/CFT Correlators}},
  \href{http://xxx.lanl.gov/abs/1102.4724}{{\tt arXiv:1102.4724}}.

\bibitem{Maldacena:2011nz}
J.~M. Maldacena and G.~L. Pimentel, {\it {On graviton non-Gaussianities during
  inflation}},  \href{http://xxx.lanl.gov/abs/1104.2846}{{\tt
  arXiv:1104.2846}}.

\bibitem{Fitzpatrick:2011ia}
A.~L. Fitzpatrick, J.~Kaplan, J.~Penedones, S.~Raju, and B.~C. van Rees, {\it
  {A Natural Language for AdS/CFT Correlators}},
  \href{http://xxx.lanl.gov/abs/1107.1499}{{\tt arXiv:1107.1499}}.

\bibitem{Paulos:2011ie}
M.~F. Paulos, {\it {Towards Feynman rules for Mellin amplitudes}},
  \href{http://xxx.lanl.gov/abs/1107.1504}{{\tt arXiv:1107.1504}}.

\bibitem{Penedones:2010ue}
J.~Penedones, {\it {Writing CFT correlation functions as AdS scattering
  amplitudes}},  \href{http://xxx.lanl.gov/abs/1011.1485}{{\tt
  arXiv:1011.1485}}.

\bibitem{Raju:2012zs}
S.~Raju, {\it {Four Point Functions of the Stress Tensor and Conserved Currents
  in AdS$_4$/CFT$_3$}},  \href{http://xxx.lanl.gov/abs/1201.6452}{{\tt
  arXiv:1201.6452}}.

\bibitem{Risager:2005vk}
K.~Risager, {\it {A Direct proof of the CSW rules}},  {\em JHEP} {\bf 0512}
  (2005) 003, [\href{http://xxx.lanl.gov/abs/hep-th/0508206}{{\tt
  hep-th/0508206}}].

\bibitem{Polchinski:1999ry}
J.~Polchinski, {\it {S matrices from AdS space-time}},
  \href{http://xxx.lanl.gov/abs/hep-th/9901076}{{\tt hep-th/9901076}}.

\bibitem{Susskind:1998vk}
L.~Susskind, {\it {Holography in the flat space limit}},
  \href{http://xxx.lanl.gov/abs/hep-th/9901079}{{\tt hep-th/9901079}}.

\bibitem{Gary:2009ae}
M.~Gary, S.~B. Giddings, and J.~Penedones, {\it {Local bulk S-matrix elements
  and CFT singularities}},  {\em Phys.Rev.} {\bf D80} (2009) 085005,
  [\href{http://xxx.lanl.gov/abs/0903.4437}{{\tt arXiv:0903.4437}}].

\bibitem{Gary:2009mi}
M.~Gary and S.~B. Giddings, {\it {The Flat space S-matrix from the AdS/CFT
  correspondence?}},  {\em Phys.Rev.} {\bf D80} (2009) 046008,
  [\href{http://xxx.lanl.gov/abs/0904.3544}{{\tt arXiv:0904.3544}}].

\bibitem{Giddings:1999qu}
S.~B. Giddings, {\it {The Boundary S matrix and the AdS to CFT dictionary}},
  {\em Phys.Rev.Lett.} {\bf 83} (1999) 2707--2710,
  [\href{http://xxx.lanl.gov/abs/hep-th/9903048}{{\tt hep-th/9903048}}].

\bibitem{Fitzpatrick:2011hu}
A.~Fitzpatrick and J.~Kaplan, {\it {Analyticity and the Holographic S-Matrix}},
   \href{http://xxx.lanl.gov/abs/1111.6972}{{\tt arXiv:1111.6972}}.

\bibitem{Osborn:1993cr}
H.~Osborn and A.~Petkou, {\it {Implications of conformal invariance in field
  theories for general dimensions}},  {\em Annals Phys.} {\bf 231} (1994)
  311--362, [\href{http://xxx.lanl.gov/abs/hep-th/9307010}{{\tt
  hep-th/9307010}}].

\bibitem{Balasubramanian:1999ri}
V.~Balasubramanian, S.~B. Giddings, and A.~E. Lawrence, {\it {What do CFTs tell
  us about anti-de Sitter spacetimes?}},  {\em JHEP} {\bf 03} (1999) 001,
  [\href{http://xxx.lanl.gov/abs/hep-th/9902052}{{\tt hep-th/9902052}}].

\bibitem{ArkaniHamed:2009si}
N.~Arkani-Hamed, F.~Cachazo, C.~Cheung, and J.~Kaplan, {\it {The S-Matrix in
  Twistor Space}},  {\em JHEP} {\bf 1003} (2010) 110,
  [\href{http://xxx.lanl.gov/abs/0903.2110}{{\tt arXiv:0903.2110}}].

\bibitem{Maldacenastrings11}
J.~Maldacena, ``Perturbative features of the wavefunction of the universe for
  pure gravity.''
\newblock Talk at Strings 2011.

\bibitem{eden1966asm}
R.~Eden, P.~Landshoff, and D.~Olive, {\em {The Analytic S-Matrix}}.
\newblock Cambridge University Press, London, 1966.

\bibitem{Vasiliev:2003ev}
M.~A. Vasiliev, {\it {Nonlinear equations for symmetric massless higher spin
  fields in (A)dS(d)}},  {\em Phys. Lett.} {\bf B567} (2003) 139--151,
  [\href{http://xxx.lanl.gov/abs/hep-th/0304049}{{\tt hep-th/0304049}}].

\bibitem{Vasiliev:1999ba}
M.~A. Vasiliev, {\it {Higher spin gauge theories: Star-product and AdS space}},
   \href{http://xxx.lanl.gov/abs/hep-th/9910096}{{\tt hep-th/9910096}}.

\bibitem{Bekaert:2005vh}
X.~Bekaert, S.~Cnockaert, C.~Iazeolla, and M.~A. Vasiliev, {\it {Nonlinear
  higher spin theories in various dimensions}},
  \href{http://xxx.lanl.gov/abs/hep-th/0503128}{{\tt hep-th/0503128}}.

\bibitem{Iazeolla:2008bp}
C.~Iazeolla, {\it {On the Algebraic Structure of Higher-Spin Field Equations
  and New Exact Solutions}},  \href{http://xxx.lanl.gov/abs/0807.0406}{{\tt
  arXiv:0807.0406}}.

\bibitem{Campoleoni:2009je}
A.~Campoleoni, {\it {Metric-like Lagrangian Formulations for Higher-Spin Fields
  of Mixed Symmetry}},  {\em Riv. Nuovo Cim.} {\bf 033} (2010) 123--253,
  [\href{http://xxx.lanl.gov/abs/0910.3155}{{\tt arXiv:0910.3155}}].

\bibitem{Boels:2008fc}
R.~Boels, K.~J. Larsen, N.~A. Obers, and M.~Vonk, {\it {MHV, CSW and BCFW:
  field theory structures in string theory amplitudes}},  {\em JHEP} {\bf 11}
  (2008) 015, [\href{http://xxx.lanl.gov/abs/0808.2598}{{\tt
  arXiv:0808.2598}}].

\bibitem{Cheung:2010vn}
C.~Cheung, D.~O'Connell, and B.~Wecht, {\it {BCFW Recursion Relations and
  String Theory}},  {\em JHEP} {\bf 1009} (2010) 052,
  [\href{http://xxx.lanl.gov/abs/1002.4674}{{\tt arXiv:1002.4674}}].

\bibitem{Boels:2010bv}
R.~H. Boels, D.~Marmiroli, and N.~A. Obers, {\it {On-shell Recursion in String
  Theory}},  {\em JHEP} {\bf 1010} (2010) 034,
  [\href{http://xxx.lanl.gov/abs/1002.5029}{{\tt arXiv:1002.5029}}].

\bibitem{Klebanov:2002ja}
I.~R. Klebanov and A.~M. Polyakov, {\it {AdS dual of the critical O(N) vector
  model}},  {\em Phys. Lett.} {\bf B550} (2002) 213--219,
  [\href{http://xxx.lanl.gov/abs/hep-th/0210114}{{\tt hep-th/0210114}}].

\bibitem{Giombi:2010vg}
S.~Giombi and X.~Yin, {\it {Higher Spins in AdS and Twistorial Holography}},
  {\em JHEP} {\bf 04} (2011) 086,
  [\href{http://xxx.lanl.gov/abs/1004.3736}{{\tt arXiv:1004.3736}}].

\bibitem{Giombi:2011ya}
S.~Giombi and X.~Yin, {\it {On Higher Spin Gauge Theory and the Critical O(N)
  Model}},  \href{http://xxx.lanl.gov/abs/1105.4011}{{\tt arXiv:1105.4011}}.

\bibitem{Gang:2010gy}
D.~Gang, Y.-t. Huang, E.~Koh, S.~Lee, and A.~E. Lipstein, {\it {Tree-level
  Recursion Relation and Dual Superconformal Symmetry of the ABJM Theory}},
  {\em JHEP} {\bf 1103} (2011) 116,
  [\href{http://xxx.lanl.gov/abs/1012.5032}{{\tt arXiv:1012.5032}}].

\bibitem{ArkaniHamed:2008yf}
N.~Arkani-Hamed and J.~Kaplan, {\it {On Tree Amplitudes in Gauge Theory and
  Gravity}},  {\em JHEP} {\bf 04} (2008) 076,
  [\href{http://xxx.lanl.gov/abs/0801.2385}{{\tt arXiv:0801.2385}}].

\end{thebibliography}\endgroup
